\documentclass[fleqn,10pt]{wlscirep}
\usepackage[utf8]{inputenc}
\usepackage[T1]{fontenc}
\usepackage{graphicx}  % needed for figures
\usepackage{dcolumn}   % needed for some tables
\usepackage{bm}        % for math
\usepackage{verbatim}   % for math
\usepackage{pdfpages}
\usepackage{epstopdf}
\usepackage{epsfig}
\usepackage{subfigure,cite}
\usepackage{xcolor}
\usepackage{amsmath}
\usepackage[most]{tcolorbox}
\usepackage{lipsum} % just for the example
\usepackage{color}
\usepackage[framemethod=TikZ]{mdframed}
\usepackage{lipsum}
\mdfdefinestyle{MyFrame}{%
	linecolor=blue,
	outerlinewidth=2pt,
	roundcorner=20pt,
	innertopmargin=\baselineskip,
	innerbottommargin=\baselineskip,
	innerrightmargin=20pt,
	innerleftmargin=20pt,
	backgroundcolor=gray!50!white}
\newcommand{\ve}[1]{\boldsymbol{#1}}
\newcommand{\te}[1]{\overline{\overline{#1}}}

\title{Programmable Nonreciprocal Metaprism}

\author[*,1]{Sajjad Taravati}
\author{George V. Eleftheriades}
\affil[1]{The Edward S. Rogers Sr. Department of Electrical and Computer Engineering, University of Toronto, Toronto, Ontario M5S 3H7, Canada.}

\affil[*]{sajjad.taravati@utoronto.ca}

\begin{abstract}
	Optical prisms are made of glass and map temporal frequencies into spatial frequencies by decomposing incident white light into its constituent colors and refract them into different directions. Conventional prisms suffer from their volumetric bulky and heavy structure and their material parameters are dictated by the Lorentz reciprocity theorem. Considering various applications of prisms in wave engineering and their growing applications in the invisible spectrum and antenna applications, there is a demand for compact apparatuses that are capable of providing prism functionality in a reconfigurable manner, with a nonreciprocal/reciprocal response. Here, we propose a nonreciprocal metasurface-based prism constituted of an array of phase- and amplitude-gradient frequency-dependent spatially variant radiating super-cells. In conventional optical prisms, nonreciprocal devices and metamaterials, the spatial decomposition and nonreciprocity functions are fixed and noneditable. Here, we present a programmable metasurface integrated with amplifiers to realize controllable nonreciprocal spatial decomposition, where each frequency component of the incident polychromatic wave can be transmitted under an arbitrary and programmable angle of transmission with a desired transmission gain. Such a polychromatic metasurface prism is constituted of frequency-dependent spatially variant transistor-based phase shifters and amplifiers for the spatial decomposition of the wave components. Interesting features include three-dimensional prism functionality with programmable angles of refraction, power amplification, and directive and diverse radiation beams. Furthermore, the metasurface prism can be digitally controlled via a field- programmable gate array (FPGA), making the metasurface a suitable solution for radars, holography applications, and wireless telecommunication systems.
\end{abstract}
\begin{document}
	
	\flushbottom
	\maketitle
	% * <john.hammersley@gmail.com> 2015-02-09T12:07:31.197Z:
	%
	%  Click the title above to edit the author information and abstract
	%
	\thispagestyle{empty}

\section{Introduction}

Optical prisms are transparent apparatuses that map temporal frequencies into spatial frequencies~\cite{Newton_1721,fork1984negative,cui2012accurate,busch20183d}. They decompose incident white light into its constituent colors and refract them into different directions. Historically, Sir Isaac Newton experimentally showed that such a phenomenon is due to the decomposition of the colors already present in the incoming light~\cite{Newton_1721}. He used a prism to show that white light is comprised of all colors in the visible spectrum and that spatial decomposition of white light is due to the inherent dispersion of glass~\cite{Newton_1721}. Over the past century, prisms have found numerous application, including interferometry~\cite{suzuki2002two}, ophthalmology~\cite{thompson1983ophthalmic}, telescopes~\cite{hensoldt1901prism,dunning1945prism}, cameras~\cite{baker1978prism,hyatt2009rotating}, microscopes and periscopes~\cite{machikhin2019modification,gorevoy2020optimization}, image observation~\cite{togino1999prism,kobayashi2002optical,takahashi2004prism}, and antenna dispersion reduction~\cite{memeletzoglou2019holey,wang2018low}. Furthermore, scientists utilize prisms to study the nature of light and human perception of light. 

Despite prisms being essential parts of optical and antenna systems, they suffer from their bulky and heavy structure, and are restricted by a reciprocal response. Considering several intriguing applications of prisms for wave engineering and their growing applications in the invisible spectrum for antenna applications, there is a demand for compact apparatuses that are capable of providing prism functionality but in a reconfigurable manner, with a nonreciprocal/reciprocal response. Electromagnetic metasurfaces have been recently proposed as two-dimensional versions of volume metamaterials and ultra compact devices for reciprocal phase front transformation of electromagnetic waves. Over the past decade, metasurfaces have seen significant advances to wave engineering in modern telecommunication and optical systems~\cite{Capasso_Sc_2011,lin2014dielectric,zheng2015metasurface,burch2017conformable,Taravati_2017_NR_Nongyro,wong2018perfect,gao2019conformal,tan2019nonreciprocal,ranaweera2019active,taravati2020full,yuan2020independent,zhang2021polarization}.

To overcome the limitations of reciprocal metasurfaces, nonreciprocal and dynamic metasurfaces have recently been proposed for extraordinary transformation of electromagnetic waves. For instance,  adding temporal variation or unilateral transistor-based circuits to conventional static metasurfaces leads to dynamic space-time metasurfaces which are capable of four-dimensional engineering of both the spatial and temporal characteristics of electromagnetic waves. Such nonreciprocal metasurfaces can be modeled by bianisotropic constitutive parameters and introduce functionalities that are far beyond the capabilities of conventional static metasurfaces. This includes nonreciprocal full-duplex wave transmission~\cite{taravati2020full}, frequency conversion~\cite{Taravati_LWA_2017,Taravati_PRB_Mixer_2018,Grbic2019serrodyne,taravati2021pure}, spatiotemporal decomposition~\cite{taravati_STMetasRev_2020,castaldi2020joint}, and space-time wave diffraction~\cite{taravati_PRApp_2019,tiukuvaara2020floquet}.

This study proposes a metasurface prism as a compact device for nonreciprocal prism-like spatial decomposition of electromagnetic waves. In conventional optical prisms, nonreciprocal devices and metamaterials, the spatial decomposition and nonreciprocity functions are fixed and noneditable. Here, we present a programmable metasurface integrated with phase- and amplitude-gradient radiating super-cells to realize controllable nonreciprocal spatial decomposition. Interesting features include a three-dimensional prism functionality with tunable angles of refraction and power amplification. The proposed polychromatic metasurface takes advantage of magnet-free nonreciprocity induced by unilateral transistors. Magnet-free nonreciprocal metasurfaces provide huge degrees of freedom for arbitrary alteration of the wavevector and temporal frequency of electromagnetic waves~\cite{Alu_PRB_2015,Taravati_PRAp_2018,Taravati_Kishk_TAP_2019,zang2019nonreciprocal_metas,Taravati_Kishk_PRB_2018,saikia2019frequency,Taravati_Kishk_MicMag_2019,wu2020space,Taravati_AMA_PRApp_2020}. Furthermore, the metasurface prism is constituted of frequency-dependent spatially variant phase shifters for spatial decomposition of wave components.

As previously mentioned, the previously reported metamaterials and metasurfaces only focus on the realization of prisms or nonreciprocal structures, which is the first step for a functional design. In addition, most of the previous prisms and nonreciprocal devices suffer from a fixed and noneditable response. Here, we present a transmissive metasurface prism to program nonreciprocal spatial decomposition. Not only the reciprocity of the metasurface prism can be reprogrammed, but also the angles of refraction of the constituent frequency components, power amplification level, and the isolation between the forward and backward transmissions can be redefined. For this purpose, the metasurface can be controlled via a low cost detachable digital sub-system integrated with unilateral amplifiers and phase shifters. We provide a proof of concept implementation where the digital control subsystem is accomplished manually. Nonetheless, the controllable and programmable features of the spatial decomposition and nonreciprocity are expected to offer completely new functionalities in conventional physical devices for applications in 5G and 6G wireless telecommunications, as well as, in optical systems.

\begin{figure*}
	\begin{center}	
		\includegraphics[width=1\columnwidth]{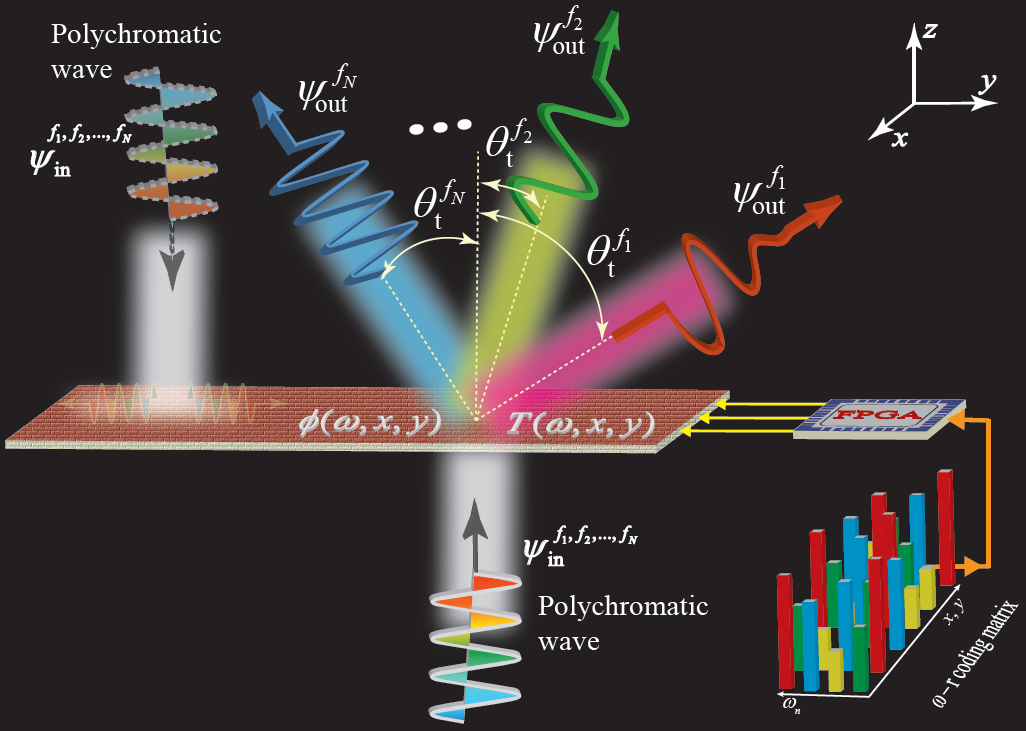}		
		\caption{Conceptual illustration of a digital programmable nonreciprocal metasurface prism. Incidence of a forward polychromatic electromagnetic (from bottom) wave leads to a prism-like full transmission to the top of the metasurface accompanied with spatial decomposition of frequency components. However, incidence of a backward polychromatic electromagnetic wave (from top) yields full absorption by the metasurface and hence no transmission to the bottom side.}
		\label{fig:1}
	\end{center}
\end{figure*}

\section{Results}

Figure~\ref{fig:1} schematically illustrates the functionality of the nonreciprocal reconfigurable metasurface prism. The metasurface is formed by an array of active frequency-dependent spatially variant inclusions providing the required  phase shift and magnitude for the forward and backward wave incidences, corresponding to the transmission from bottom to top and from top to bottom, respectively. In general, the metasurface provides the frequency-dependent spatially variant phase shift of $\phi^{\text{F}}(\omega,x,y)$ for the forward wave transmission(shown in Fig.~\ref{fig:1}), which is different than the one of the phase shift provided for the backward direction (shown in the left side of Fig.~\ref{fig:1}), i.e., $\phi^{\text{B}}(\omega,x,y) \neq \phi^{\text{F}}(\omega,x,y)$. In addition to the phase shift difference provided by the metasurface for the forward and backward directions, the metasurface exhibits a frequency-dependent spatially variant transmission gain of $T^{\text{F}}(\omega,x,y)$ for the forward wave transmission, whereas it provides a transmission loss for the backward direction  i.e., $T^{\text{F}}(\omega,x,y)  >1 \gg T^{\text{B}}(\omega,x,y)$.

Assume the polychromatic electromagnetic wave $\ve{\psi}_\text{in}$ comprising different frequency components, i.e., $\omega_1,\omega_2,...,\omega_\text{N}$ is traveling along the $+z$ direction and passes through the metasurface (Fig.~\ref{fig:1}). As the metasurface introduces frequency-dependent spatially variant phase shifts in the forward direction, the frequency components of the incident polychromatic wave acquire different phase shifts at different locations on the $x-y$ plane, $\phi_n(x,y)$. In general, the phase and magnitude profiles of the metasurface, $\phi^{\text{F,B}}(\omega,x,y)$ and $T^{\text{F,B}}(\omega,x,y)$, are designed such that each transmitted frequency component $\ve{\psi}_\text{out}^{f_n}$ is transmitted under a desired transmission angle $\theta_\text{t}^{f_n}$ with a specified transmission gain $|\ve{\psi}_{\text{out},n}|$. 

Assuming a constant gradient phase shift along the metasurface, the generalized Snell’s law of refraction yields
\begin{equation}\label{eq:theta}
\frac{\partial \phi_{\text{x},n}}{\partial x}=k_n \left[\sin(\theta_\text{x}^\text{trns}) -\sin(\theta_\text{x}^\text{inc}) \right],
\end{equation}
\begin{equation}\label{eq:phi}
\frac{\partial \phi_{\text{y},n}}{\partial y}=k_n \left[\sin(\theta_\text{y}^\text{trns}) -\sin(\theta_\text{y}^\text{inc}) \right],
\end{equation}
for the reception state. Here, $k_\text{1}$ and $k_\text{2}$ are the wave numbers in region 1 and region 2, respectively. Considering a constant phase gradient $\partial \phi_\text{MS}/\partial x$, the outgoing wave acquires anomalous refraction with respect to the incident
wave, whereas a spatially variant gradient, i.e, $\partial \phi_\text{MS}/\partial x$, leads to arbitrary radiation beams which enables beam-forming and advanced beam steering applications.

To analyze the metasurface prism, we start from the general case of a bianisotropic temporally dispersive spatially variant metasurface characterized by the spectral relations
\begin{subequations}
	\label{Eq:ConstRel}
	\begin{equation}
	\ve{D}(\omega,\textbf{r}) = \te{\epsilon}(\omega,\textbf{r}) \cdot \ve{E}(\omega,\textbf{r}) + \te{\xi}(\omega,\textbf{r}) \cdot\ve{H}(\omega,\textbf{r}),
	\end{equation}
	\begin{equation}
	\ve{B}(\omega,\textbf{r}) =  \te{\zeta}(\omega,\textbf{r}) \cdot \ve{E}(\omega,\textbf{r}) + \te{\mu}(\omega,\textbf{r}) \cdot\ve{H}(\omega,\textbf{r}).
	\end{equation}
\end{subequations}

Following the GSTC method~\cite{kuester2003averaged,idemen2011discontinuities} for analysis of zero-thickness metasurfaces, the metasurface will be characterized by the continuity equations, i.e.,
\begin{subequations}
	\label{Eq:MSBC}
	\begin{align}
	\hat{z}&\times\Delta\ve{H}(\omega,\textbf{r})\\ \nonumber
	&=j\omega\epsilon_0\te{\chi}_\text{ee}(\omega,\textbf{r}) \cdot\ve{E}_\text{av}(\omega) +jk \te{\chi}_\text{em}(\omega,\textbf{r}) \cdot\ve{H}_\text{av}(\omega),
	\end{align}
	\begin{align}
	\Delta&\ve{E}(\omega,\textbf{r}) \times\hat{z}\\ \nonumber
	&=j\omega\mu_0 \te{\chi}_\text{mm}(\omega,\textbf{r}) \cdot\ve{H}_\text{av}(\omega)+jk \te{\chi}_\text{me}(\omega,\textbf{r})\cdot\ve{E}_\text{av}(\omega),
	\end{align}
\end{subequations}
Equation~\eqref{Eq:MSBC} provides the relation between the electromagnetic fields on both sides of the metasurface to its susceptibilities, while assuming no normal susceptibility components. Here, $\Delta$ and the subscript `av' denote the difference of the fields and the average of the fields between both sides of the metasurface.
The susceptibilities in~\eqref{Eq:MSBC} are related to the constitutive parameters in~\eqref{Eq:ConstRel} as
\begin{subequations}
	\label{eq:const_para_xi}
	\begin{equation}
	\te{\epsilon}(\omega,\textbf{r})=\epsilon_0(\te{I}+\te{\chi}_\text{ee}(\omega,\textbf{r})),\quad\te{\mu}(\omega,\textbf{r})=\mu_0(\te{I}+\te{\chi}_\text{mm}(\omega,\textbf{r})),
	\end{equation}
	\begin{equation}
	\te{\xi}(\omega,\textbf{r})=\te{\chi}_\text{em}(\omega,\textbf{r})/c_0,\quad\te{\zeta}(\omega,\textbf{r})=\te{\chi}_\text{me}(\omega,\textbf{r})/c_0.
	\end{equation}
\end{subequations}
We shall then find the susceptibilities providing the nonreciprocal response for the metasurface. This consists in substituting the electromagnetic fields of the corresponding transformation into~\eqref{Eq:MSBC}. Specifically, the transformation consists in passing an incident forward $+z$-propagating plane wave through the metasurface with transmission coefficient $T>>1$ and absorbing an incident backward $-z$-propagating plane wave with transmission coefficient $T<<1$. The result reads

\begin{subequations}
	\label{eq:ChiTrans}
	\begin{equation}
	\label{eq:ChiTrans1}
	\te{\chi}_\text{ee}(\omega,\textbf{r})= -j\frac{ v(\textbf{r})}{\omega} \begin{pmatrix} 1 & 0 \\ 0 & 1 \end{pmatrix},
	\end{equation}
	\begin{equation}
	\te{\chi}_\text{mm}(\omega,\textbf{r})= -j\frac{ v(\textbf{r})}{\omega} \begin{pmatrix} 1 & 0 \\ 0 & 1 \end{pmatrix},
	\end{equation}
	\begin{equation}
	\label{eq:ChiTrans2}
	\te{\chi}_\text{em}(\omega,\textbf{r})= j\frac{ v(\textbf{r})}{\omega} \begin{pmatrix} 0 & 1 \\ -1 & 0 \end{pmatrix},
	\end{equation}
	\begin{equation}
	\te{\chi}_\text{me}(\omega,\textbf{r})= j\frac{ v(\textbf{r})}{\omega} \begin{pmatrix} 0 & -1 \\ 1 & 0 \end{pmatrix},
	\end{equation}
\end{subequations}
showing that nonreciprocal spatial decomposition in the metasurface is due to the electric-magnetic coupling contributions, $\te{\chi}_\text{em}(\omega,\textbf{r})$ and $\te{\chi}_\text{me}(\omega,\textbf{r})$.

\begin{figure}
	\begin{center}	
		\includegraphics[width=0.8\columnwidth]{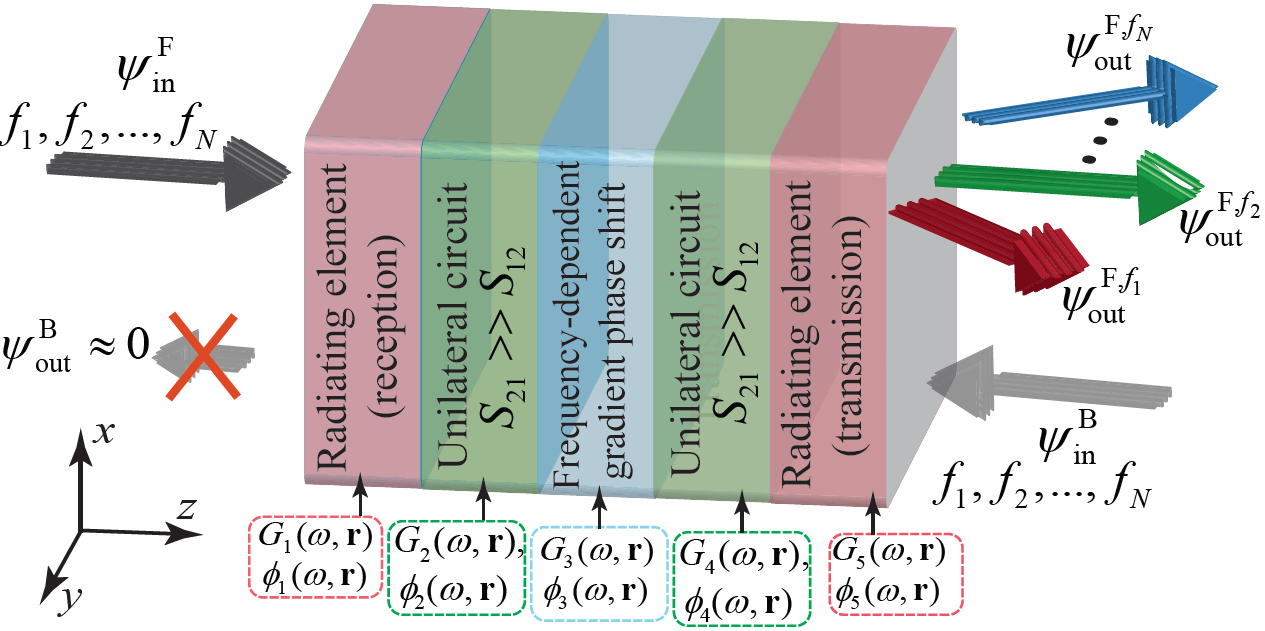}	
		\caption{Composition of the programmable nonreciprocal metasurface prism. The metasurface operation is governed by a five-stage architecture, each stage proving the required transmission phase and magnitude required for a controllable and programmable nonreciprocal prism functionality.}
		\label{fig:oper}
	\end{center}
\end{figure}

Figure~\ref{fig:oper} shows the operation principle of the metasurface prism constituted of five distinct operations. These five operations are accomplished through five different electronic and electromagnetic components, proving a strong flexibility and leverage for achieving the required programmable and controllable prism functionality as well as nonreciprocal transmission. The incoming polychromatic electromagnetic wave from the left is received by the radiating antenna elements characterized by the transmission loss of $G_1(\omega,\textbf{r})$ and transmission phase of $\phi_1(\omega,\textbf{r})$. Next, the received signal by the antenna elements passes through the unilateral circuit characterized by the transmission gain of $G_2(\omega,\textbf{r})$ and transmission phase of $\phi_2(\omega,\textbf{r})$, introducing the required power amplification and phase shift in the forward direction and a desired transmission loss in the reverse direction. Then, in the third stage, the signal enters into a phase shifter characterized by the transmission loss of $G_3(\omega,\textbf{r})$ and transmission phase of $\phi_3(\omega,\textbf{r})$ providing a desired gradient or nonlinear phase shift. Next, the signal experiences another round of controlled frequency-dependent spatially-varying amplification and phase shift by the second unilateral circuit characterized by the transmission gain of $G_4(\omega,\textbf{r})$ and transmission phase of $\phi_4(\omega,\textbf{r})$. Finally, the processed polychromatic wave is reradiated to the right side of the metasurface prism using another radiating antenna element characterized by the transmission loss of $G_5(\omega,\textbf{r})$ and transmission phase of $\phi_5(\omega,\textbf{r})$. Here we aim to design each of these five steps, corresponding to five phase shifts and transmission loss/gain so that each frequency component of the polychromatic wave acquires a desired phase (and amplitude). As a result, due to the phase difference between the frequency components at the output of the metasurface in the forward direction, in the right side of the metasurface, each frequency component of the polychromatic wave is transmitted under a specific angle of transmission. In contrast to the forward direction, the incoming wave in the backward direction from the right side cannot pass through the metasurface due to the absorption and reflection provided by two unilateral devises.

In general, the acquired phase and power gain for each frequency component reads 
\begin{subequations}
	\begin{equation}
	\phi_n(\textbf{r})=\sum_{m=1}^5  \phi_{m}(\omega_n,\textbf{r}),
	\end{equation}
	\begin{equation}
	T_n(\textbf{r})=\sum_{m=1}^5 G_{m}(\omega_n,\textbf{r}).
	\end{equation}
\end{subequations}

\begin{figure}
	\begin{center}
		\includegraphics[width=0.8\columnwidth]{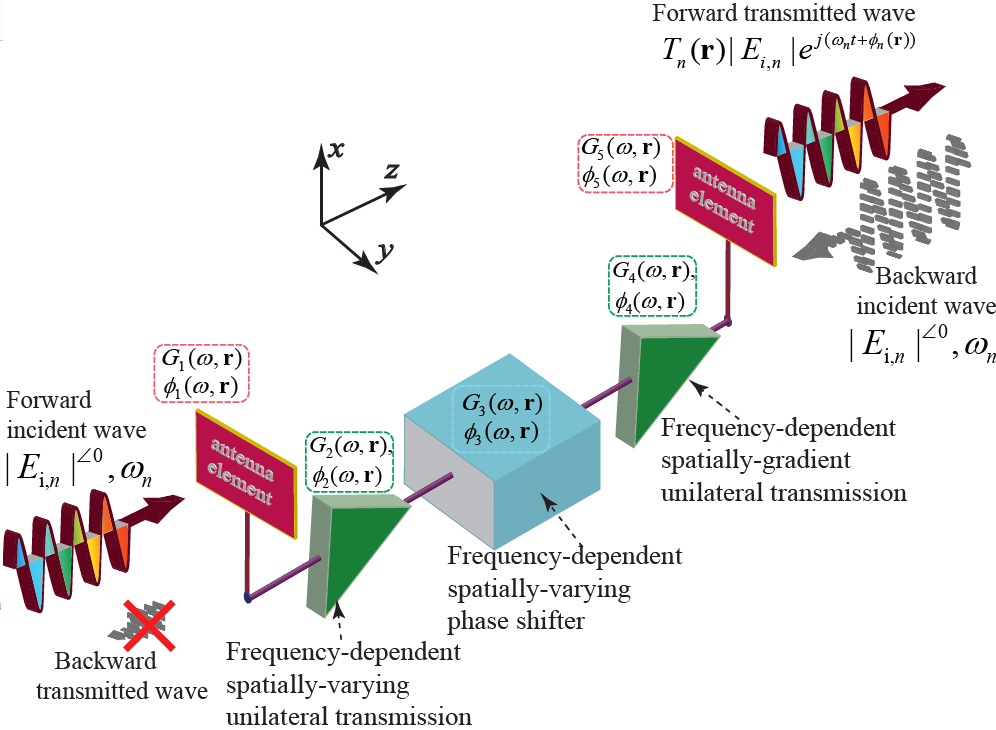}	
		\caption{Architecture of a radiating super-cell incorporating frequency-dependent spatially-varying passive and active electronic-electromagnetic elements.}
		\label{fig:UC}
	\end{center}
\end{figure}

Figure~\ref{fig:UC} shows the case of the wave incidence to a radiating super-cell incorporating frequency-dependent spatially-varying passive and active electronic-electromagnetic elements. Two microstrip patch antenna elements are interconnected through two unilateral transistor-based amplifiers and a phase shifter. In general, all five elements, i.e., two patch antennas, two unilateral transistor-based amplifiers and the phase shifter have the capability to be controlled and programmed in a way that a diverse set of responses is achievable, including three-dimensional scanning with arbitrary transmission angle and magnitude, as well as isolation between the forward and backward transmitted waves.
Considering a plane wave, expressed as $E_{\text{i},n} \exp{(j\omega_n t)}$, impinging to the left patch antenna element shown in the left side of Fig.~\ref{fig:UC} in the forward direction. Then, the transmitted wave in the right side of the super-cell in Fig.~\ref{fig:UC} reads $T_n(\textbf{r}) E_{\text{i},n} \exp{(j[\omega_n t+\phi_n (\textbf{r})])}$. 

\begin{figure*}
	\begin{center}
		\includegraphics[width=1\columnwidth]{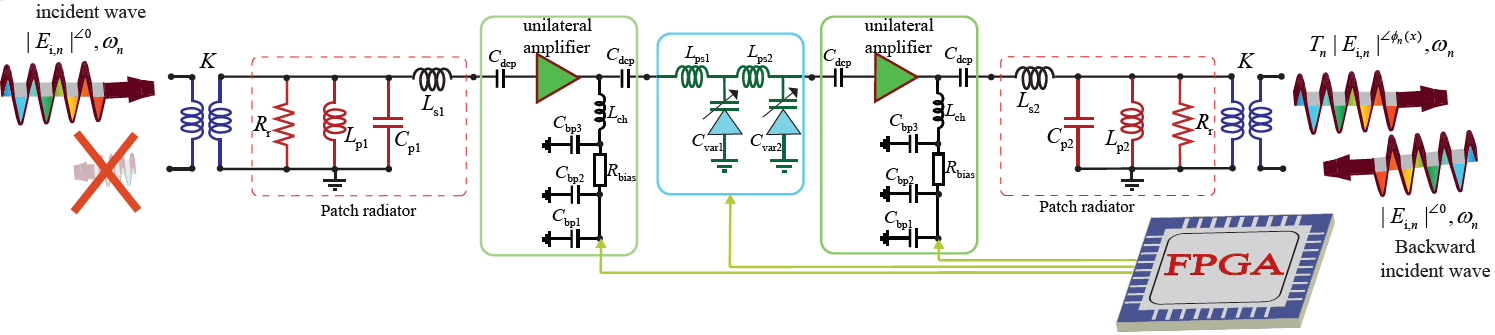}	
		\caption{A circuit model for the five-stage frequency-dependent spatially-varying radiating super-cell in Fig.~\ref{fig:UC}.}
		\label{fig:UC2}
	\end{center}
\end{figure*}

Figure~\ref{fig:UC2} illustrates a schematic of the circuit of the super-cell. This figure provides the details of the circuit of the unilateral transistor-based amplifier. The phase shift and transmission power gain provided by the super-cell is governed by a coding signal provided through an FPGA to different parts of the super-cell. An equivalent circuit model is presented for the patch antenna elements that includes a radiation resistance $R_\text{r}$, and a parallel capacitance $C_\text{p}$, a parallel inductance $L_\text{p}$ and a series inductance $L_\text{s}$. Here, the coupling of the incident wave in the free-space to the super-cell in the left and the coupling of the transmitted wave from super-cell to the free-space in the right are modeled by two transformers. The circuit of the unilateral transistor-based amplifier is composed of five capacitors, one inductance and one resistor. Two decoupling capacitances, both denoted by $C_\text{dcp}$, inhibit leakage of the DC bias signal to the RF path of the super-cell, while three by-pass capacitances, $C_\text{bp1}$, $C_\text{bp2}$ and $C_\text{bp3}$, short-circuit different undesired radio-frequency signals in the DC bias line, and a choke inductance $L_\text{ch}$ guarantees no leakage of the radio-frequency signal to the DC line. In addition, a series resistance $R_\text{bias}$ is considered for controlling the level of the flowing DC current of the amplifier. The circuit of the gradient phase shifter is modeled by two inductances, $L_\text{ps1}$ and $L_\text{ps2}$, and two variable capacitances, $C_\text{var1}$ and $C_\text{var2}$. It should be noted that, depending on the frequency of the operation of the metasurface, different gradient phase shifters may be utilized. For instance, in sub-microwave frequencies the phase shifter may be constructed by lumped variable capacitances and lumped inductances. However, at higher frequencies, the gradient phase shifter can be formed by distributed transmission-line-based phase shifters controlled by varactors.

Once the super-cells are designed, we can then construct the complete metasurfaces which is formed by an array of phase-(and/or amplitude-) gradient super-cells, where the distance between the center-to-center of two adjacent super-cells in the $x-y$ plane is denoted by $d$ with a nominal value of half-wavelength. Here, we form discrete phase and amplitude profiles $\phi_k(x,y,\omega)$ and $T_k(x,y,\omega)$, as 
\begin{equation}\label{eq:phi_tx}
\phi(kd,\omega)=\phi_k(\omega), \qquad \text{and} \qquad T(kd,\omega)=T_k(\omega).
\end{equation}

\begin{figure}
	\begin{center}
		\includegraphics[width=0.7\columnwidth]{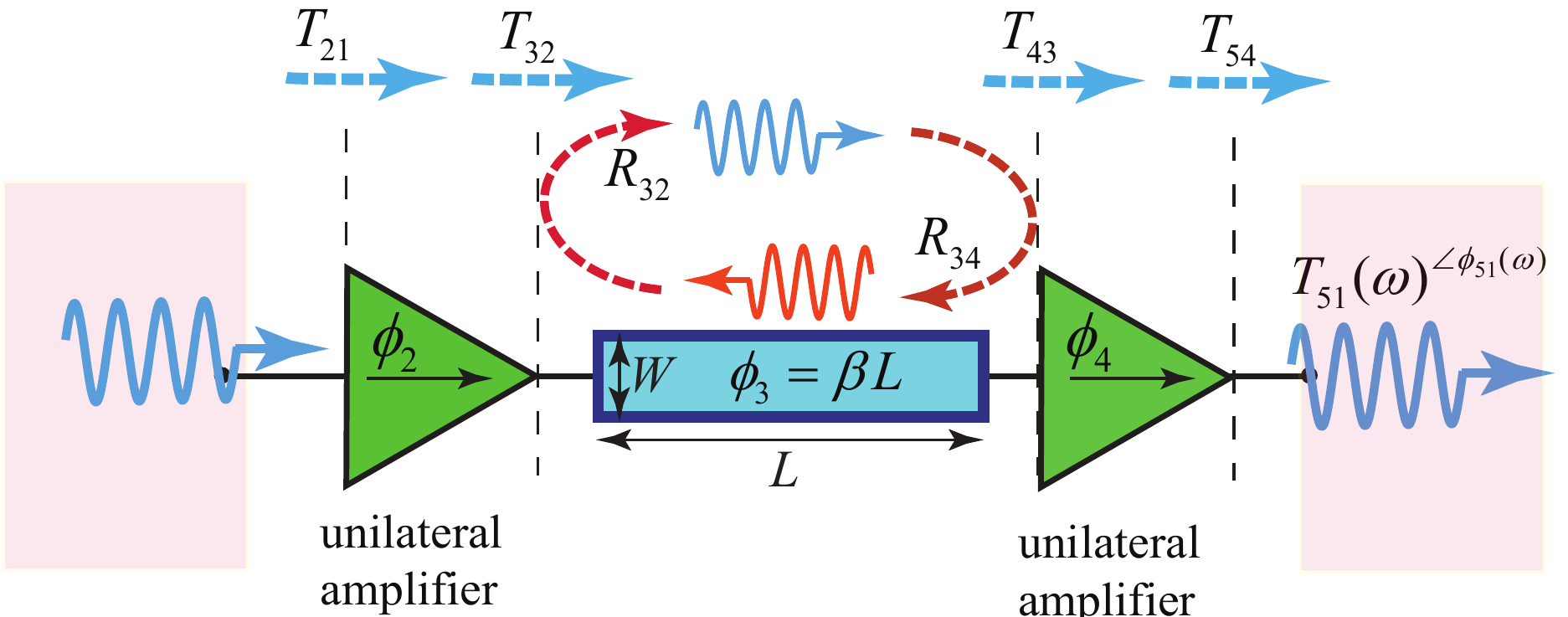}	
		\caption{Illustrative schematic of the multiple reflection and transmission inside the super-cell. By controlling the transmission phase shift and magnitude of the super-cell, arbitrary phase shifts and transmission magnitudes can be achieved at a given frequency across the frequency band via different parameters of the unilateral amplifiers and central phase shifters.}
		\label{fig:sch_tune}
	\end{center}
\end{figure}

Figure~\ref{fig:sch_tune} shows the wave transmission through the super-cell, where the $S_{11}$, $S_{21}$, $S_{12}$, and $S_{22}$, as well as width and length of the central transmission-line-based phase shifter directly affect the transmission phase (and the amplitude). As a result, by controlling the aforementioned parameters, a versatile control over the transmission phase and amplitude of each super-cell can be achieved. It should be noted that the backward wave, propagating along the $-z$ direction inside the middle transmission-line-based phase shifter, is due to reflection at the interface between the central phase shifter and second amplifier, denoted by $R_{34}$. We shall next apply the boundary conditions at the interface between regions $m$ and $m+1$ and then determine the total transmission from the input of the super-cell to its output. The total transmission and total reflection coefficients between regions $m$ and $m+1$ read~\cite{Chew_1995}

\begin{subequations}
	\begin{equation}
	\widetilde{T}_{m+1,m}(\omega) =\frac{T_{m+1,m}(\omega) e^{-j (\beta_{m}-\beta_{m+1}) z}}{1-R_{m+1,m}(\omega) \widetilde{R}_{m+1,m+2}(\omega) e^{-j 2\beta_{m+1} d_{m+1}}},
	\label{eqa:xxx}
	\end{equation}
	\begin{equation}
	\widetilde{R}_{m,m+1}(\omega) = \frac{R_{m,m+1}(\omega)+\widetilde{R}_{m+1,m+2}(\omega) e^{-j 2\beta_{m+1} d_{m+1}}}{1+R_{m,m+1}(\omega) \widetilde{R}_{m+1,m+2}(\omega) e^{-j 2\beta_{m+1} d_{m+1}}}.
	\label{eqa:www}
	\end{equation}
	\label{eqa:TRtotal}
\end{subequations}

Here, $R_{m,m+1}$ is the local reflection coefficient within region $n$ between regions $n$ and $m+1$. The local transmission coefficient from region $m$ to region $m+1$ is then found as $T_{m+1,m}=1+R_{m,m+1}$. Then, the total transmission coefficient through the super-cell of Fig.~\ref{fig:sch_tune} reads 

\begin{equation}
T_\text{51}(\omega)=\prod_{m=1}^{4} \widetilde{T}_{m+1,m}(\omega) e^{-j \beta_m d_m},
\label{eqa:mm2}
\end{equation}

\noindent where $\widetilde{T}_{m+1,m}$, for $m=1,\ldots,4$ is given in~\eqref{eqa:xxx} with~\eqref{eqa:www}. Equation~\eqref{eqa:mm2} shows that by controlling different parameters of the super-cell, one may achieve arbitrary phase shifts and transmission gain. The total forward and backward wave transmissions, $S_{21}(\omega)$ and $S_{12}(\omega)$, through the super-cell in Fig.~\ref{fig:sch_tune} may be then expressed as
	\begin{subequations}
		\begin{equation}
		S_{21}(\omega)=T_\text{51}(\omega)= \left(1-R_\text{12}(\omega) \right)
		\frac{S_\text{21}^\text{Amp1}(\omega)}{1-R_{32}(\omega) R_{34}(\omega) e^{-j 2\beta_{3} L}} (1-R_\text{34}(\omega) ) S_\text{21}^\text{Amp2}(\omega),
		\label{eqa:mm3}
		\end{equation}
		\begin{equation}
		S_{12}(\omega)=T_\text{15}(\omega)= \left(1-R_\text{45}(\omega) \right)
		\frac{S_\text{12}^\text{Amp2}(\omega)}{1-R_{34}(\omega) R_{32}(\omega) e^{-j 2\beta_{3} L}} (1-R_\text{32}(\omega) ) S_\text{12}^\text{Amp1}(\omega)
		\label{eqa:mm4}
		\end{equation}
	\end{subequations}
	where $\widetilde{R}_{34}(\omega)=R_{34}(\omega)$. Given the fact that the forward transmission of transistor-based amplifiers is much larger than their backward transmission, i.e., $S_\text{21}^\text{Amp1}(\omega)>1 >>S_\text{12}^\text{Amp1}(\omega)\approx0$ and 
	$S_\text{21}^\text{Amp2}(\omega)>1 >>S_\text{12}^\text{Amp2}(\omega)\approx0$, then one can show that the forward wave transmission is much larger than the backward wave transmission, $|S_{21}(\omega)|>>|S_{12}(\omega)|$. The total power gain of the metasurface is computed based on the average gain of the supercells~\cite{gatti2004computation}
	\begin{equation}\label{eq:11}
	|S_\text{21,tot}(\omega)|^2=10 \log \dfrac{P_\text{out}(\omega)}{P_\text{in}}=20 \log \left(\sum_{n=1}^N |S_{21,n}(\omega)|\right)-20 \log N,
	\end{equation}
where $N$ is the number of the super-cells of the metasurface and $S_{21,n}$ is the total transmission of the $n$th super-cell.

\begin{figure}
	\begin{center}
		\subfigure[]{\label{fig:203}
			\includegraphics[width=0.33\columnwidth]{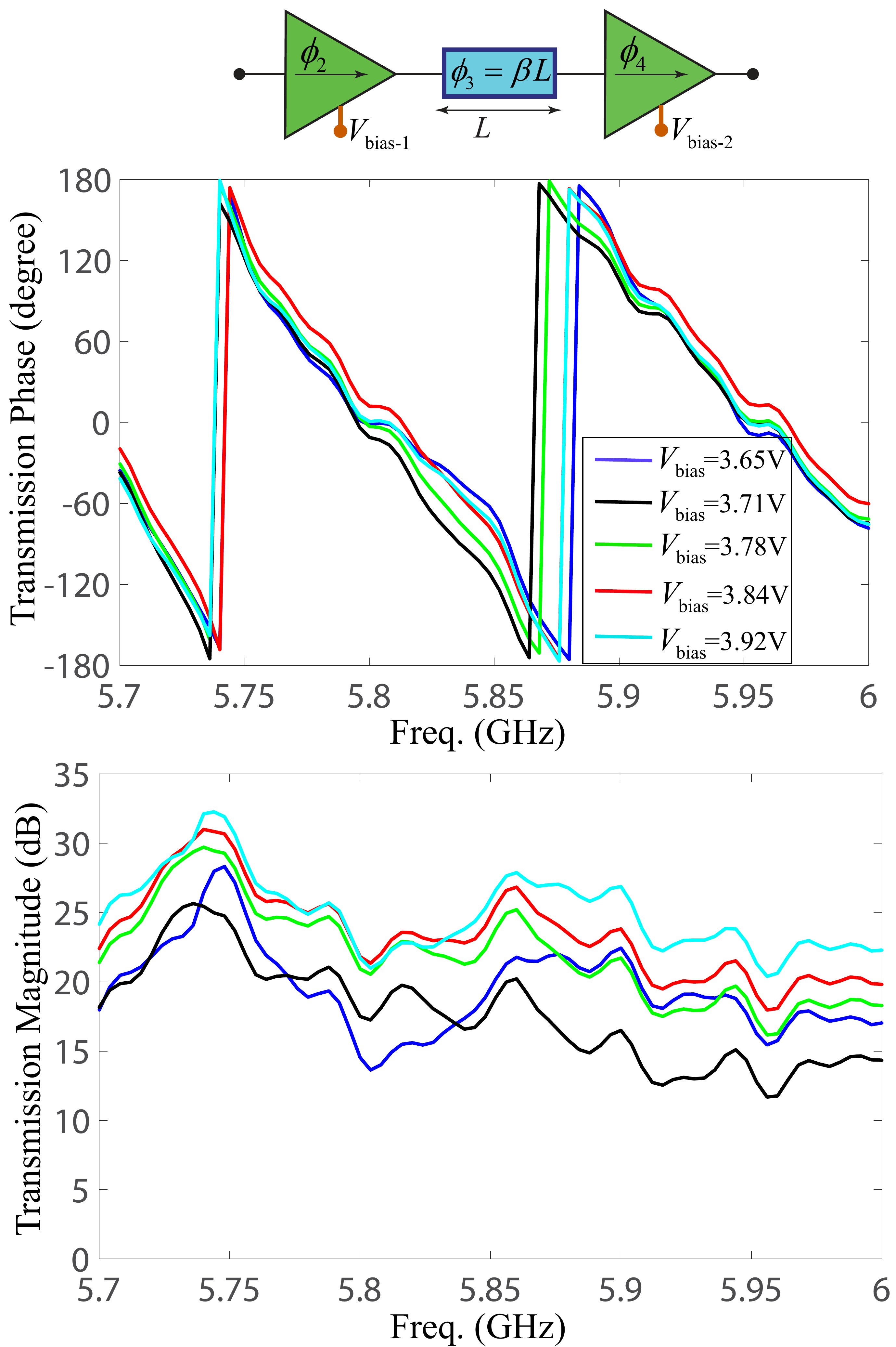}}
		\subfigure[]{\label{fig:475}
			\includegraphics[width=0.33\columnwidth]{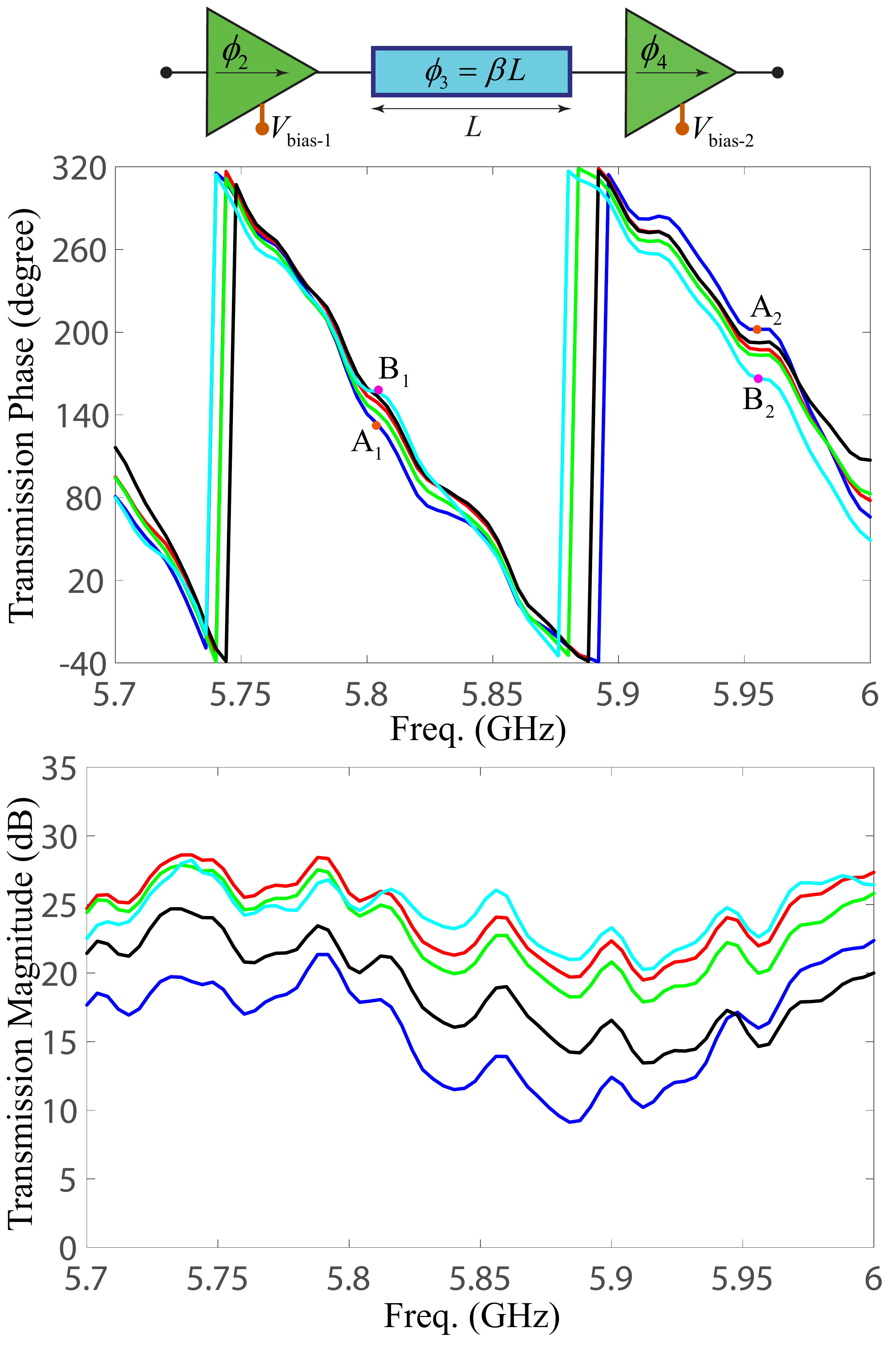}}\\	
		\subfigure[]{\label{fig:679}
			\includegraphics[width=0.33\columnwidth]{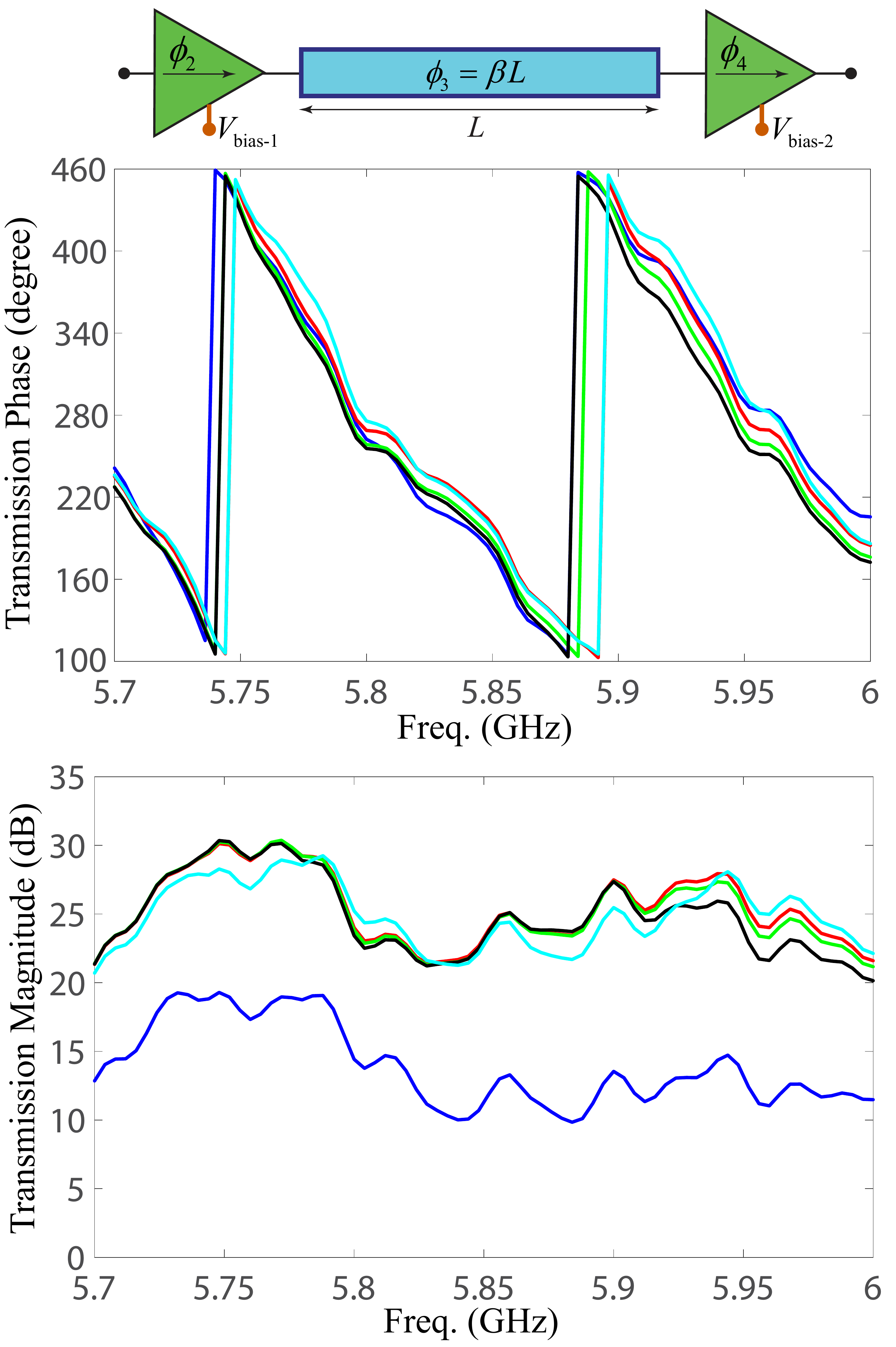}}	
		\subfigure[]{\label{fig:950}
			\includegraphics[width=0.33\columnwidth]{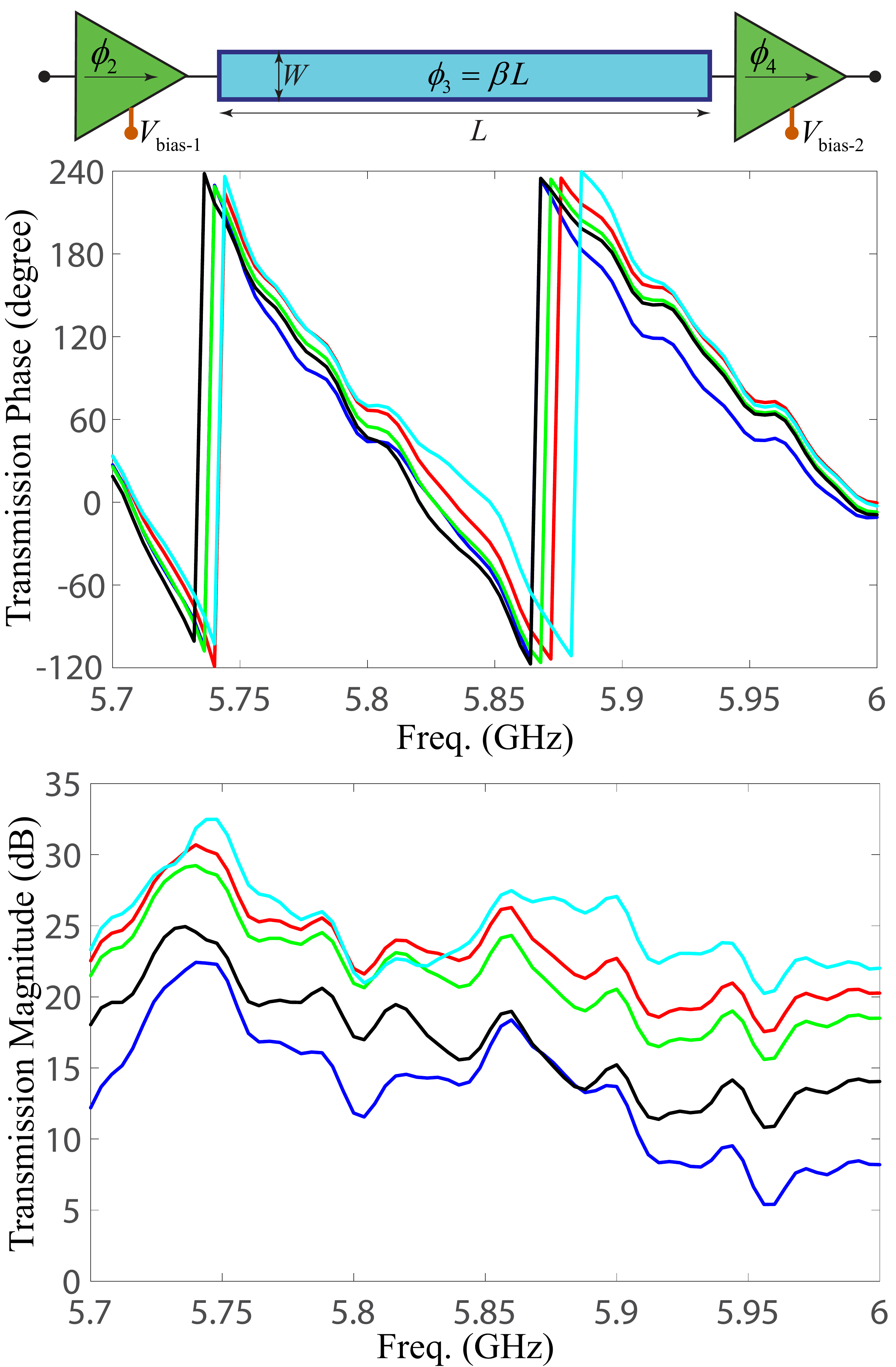}}			
		\caption{Experimental results for controlling the transmission phase shift (and magnitude) of the super-cell via unilateral amplifiers and central phase shifters for $V_\text{bias-1}=V_\text{bias-2}=V_\text{bias}$. (a) $L=203$ mils. (b) $L=475$ mils. (c) $L=679$ mils. (d) $L=950$ mils.}
		\label{fig:gradient}
	\end{center}
\end{figure}

Figures~\ref{fig:203} to~\ref{fig:950} plot the experimental results for the total phase and magnitude transmissions through the super-cell across the frequency band. These figures demonstrate the effect of the DC bias, $V_\text{bias}$, of the unilateral transistor-based amplifiers and the length of the central phase shifter, $L$, on the transmission phase (and amplitude). For instance, Fig.~\ref{fig:475} highlights the phase difference for two curves corresponding to the $V_\text{bias}=3.65 $V and $V_\text{bias}=3.92$V at two frequencies 5.8~GHz and 5.95~GHz, shown by A$_1$, B$_1$, A$_2$ and B$_2$. As can be seen, the achieved phase shift for $V_\text{bias}=3.65$V at 5.8~GHz is $23$ degrees greater than the one for $V_\text{bias}=3.92$V at 5.8~GHz, whereas the achieved phase shift for $V_\text{bias}=3.65$V at 5.95~GHz is $33$ degrees lower than the one for $V_\text{bias}=3.92$V at 5.95~GHz. As a result, one can achieve arbitrary transmission phase shifts at each super-cell simply by controlling the DC bias of the amplifiers as well as the phase shift of the central phase shifter through the DC bias of the varactor. Hence, various and different sets of linear/nonlinear phase profiles can be achieved at different frequencies across the frequency band via an FPGA. This provides the opportunity to achieve arbitrary angles and amplitudes of transmission for each frequency component. In this proof of concept we control the functionality and DC bias of transistors manually. However, 
as the prism is compatible with digital controlling systems, e.g., FPGA, integration of the digital sub-circuit can be accomplished in commercialized version of the metasurface through a standard engineering approach.

\begin{figure*}
	\includegraphics[width=1\columnwidth]{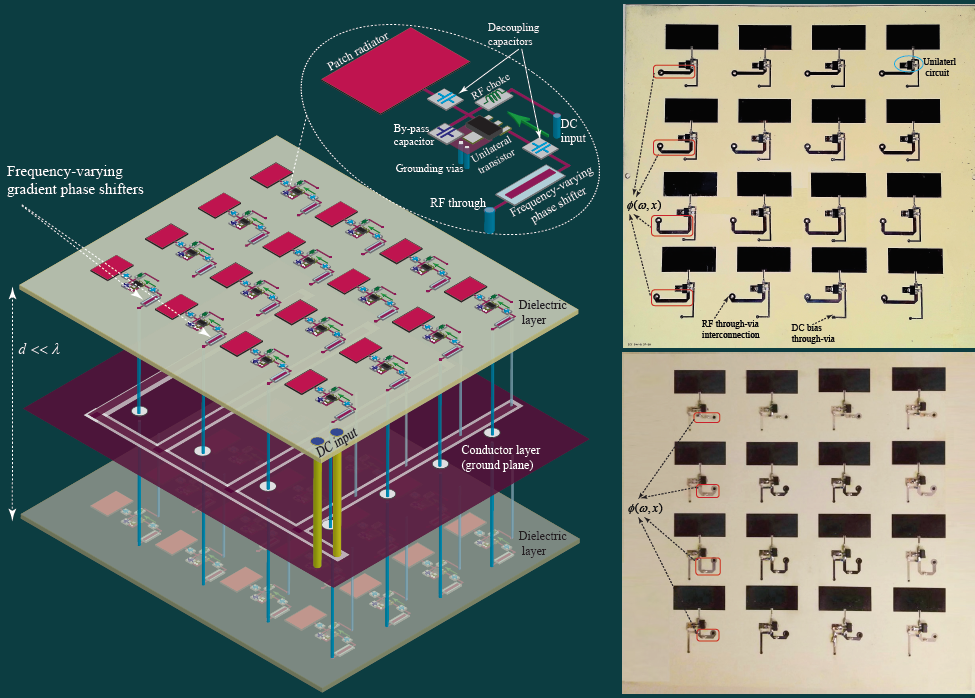}	
	\caption{Nonreciprocal Polychromatic metasurface. (a) An exploded-view of the architecture of the polychromatic transmissive metasurface. (b) Top and bottom layers of the fabricated metasurface.}
	\label{fig:arch}
\end{figure*}

\begin{figure}
	\begin{center}
		\subfigure[]{\label{fig:NF_photoL}
			\includegraphics[width=0.46\columnwidth]{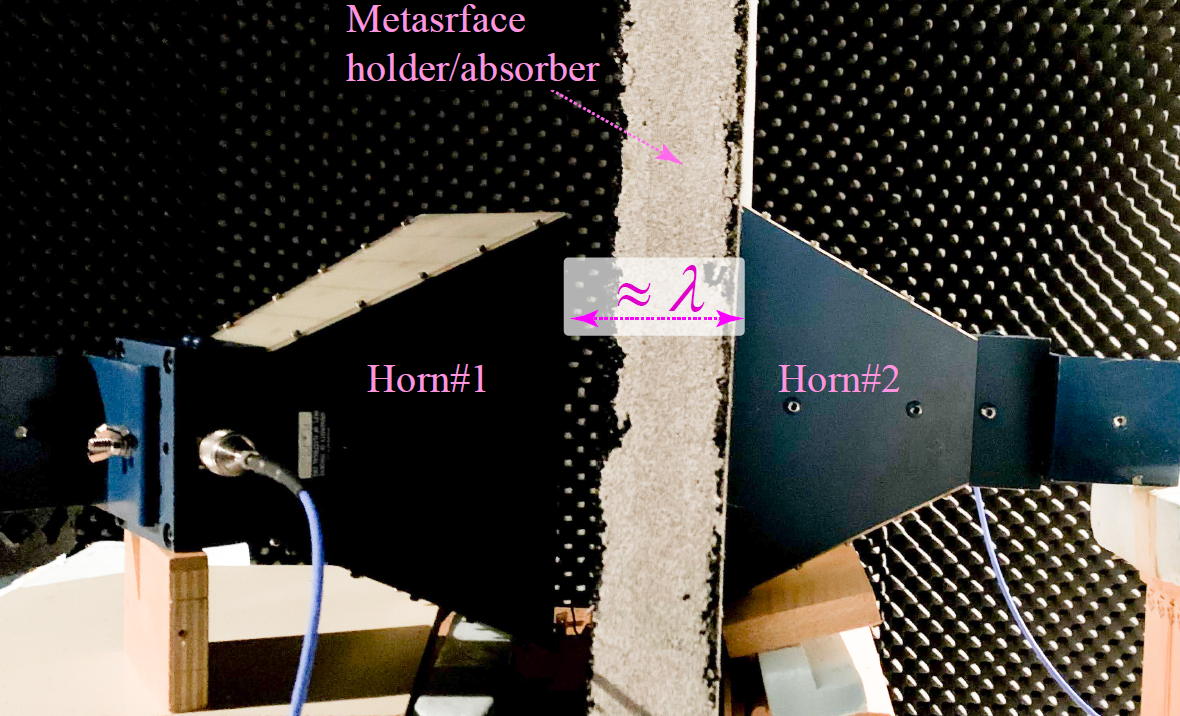}}	
		\subfigure[]{\label{fig:NF_beam}
			\includegraphics[width=0.45\columnwidth]{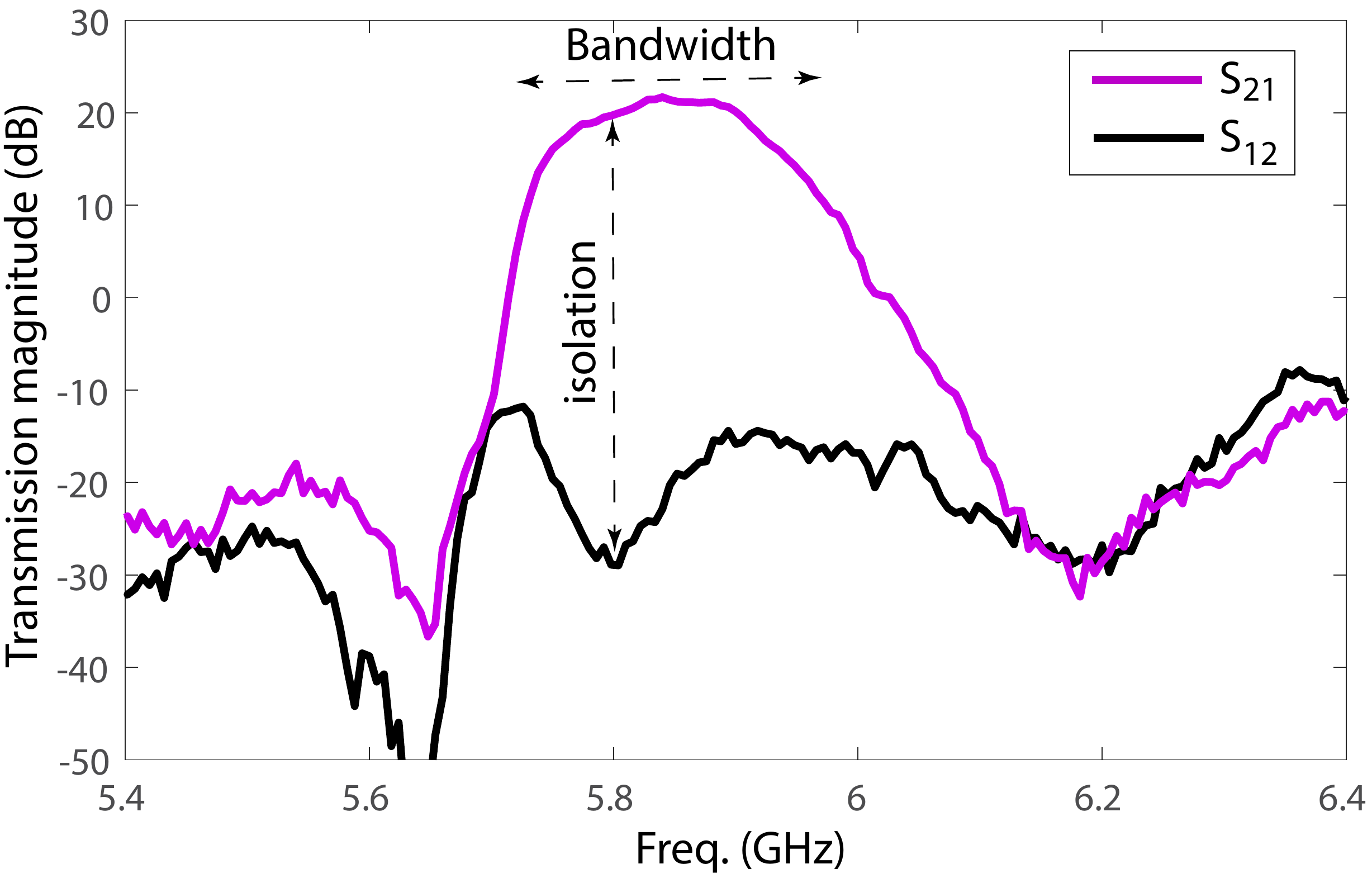}}		
		\caption{Near-field experiment. (a) An image of the near-field experimental set-up. (b) Near-field frequency response, showing a fractional frequency bandwidth of $4.3\%$, i.e., from 5.73 GHz to 5.98 GHz.}
		\label{fig:NF}
	\end{center}
\end{figure}

Figure~\ref{fig:arch}(a) illustrates an exploded schematic of the $4\times 4$ programmable nonreciprocal metasurface prism. The nonreciprocal metasurface prism is designed to operate in the frequency range from 5.75 to 6~GHz, possessing a deeply subwavelength thickness, specifically $\delta\approx\lambda_0/28$, where $\lambda_0$ is the wavelength at the center frequency, $5.875$~GHz, of the operating frequency range. Figure~\ref{fig:arch}(b) shows the top and bottom layers of the fabricated metasurface prism. We first evaluate the near-field performance of the metasurface prism, specifically its transmission gain and loss in the forward and backward directions. This is because one of the main applications of such a metasurface would be to place it in front and very close to a source antenna to achieve a required gain, nonreciprocity or prism functionality. Figure~\ref{fig:NF_photoL} shows a photo of the near-field measurement set-up. The measurements are carried out by an E8361C Agilent vector network analyzer where two horn antennas are placed at the two sides of the metasurface to transmit and receive the electromagnetic wave. Figure~\ref{fig:NF_beam} plots the near-field experimental results for the transmission scattering parameters of the metasurface prism versus frequency for normally aligned transmit and receive horn antennas. In the forward direction, more than $20$~dB transmission gain is achieved in the frequency range of interest, while in the backward direction, more than $20$~dB transmission loss is achieved across the same frequency range, corresponding to an isolation of more than $40$~dB.

\begin{figure}
	\begin{center}
		\includegraphics[width=0.55\columnwidth]{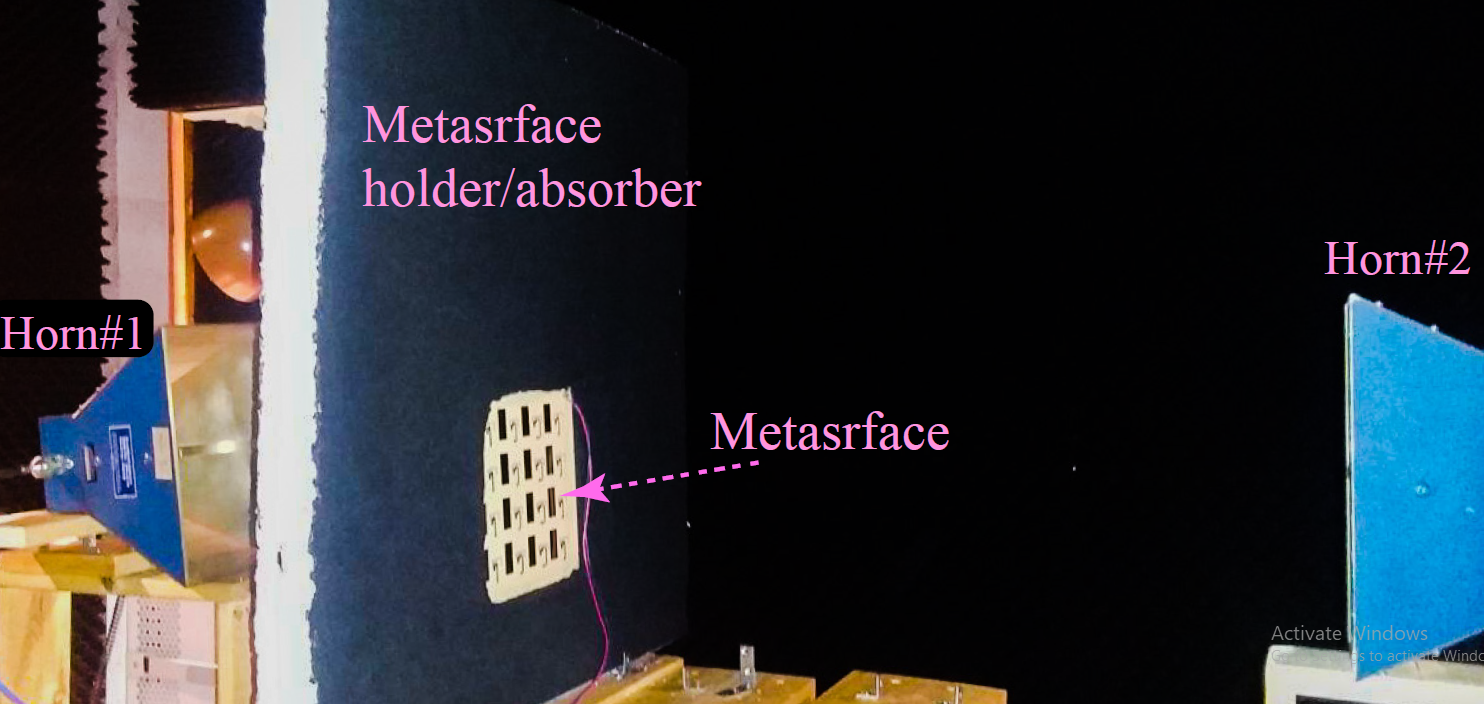}	
		\caption{An image of the far-field experimental set-up.}
		\label{fig:meas_su}
	\end{center}
\end{figure}

\begin{figure*}
	\begin{center}	
		\subfigure[]{\label{fig:s1}
			\includegraphics[width=0.32\columnwidth]{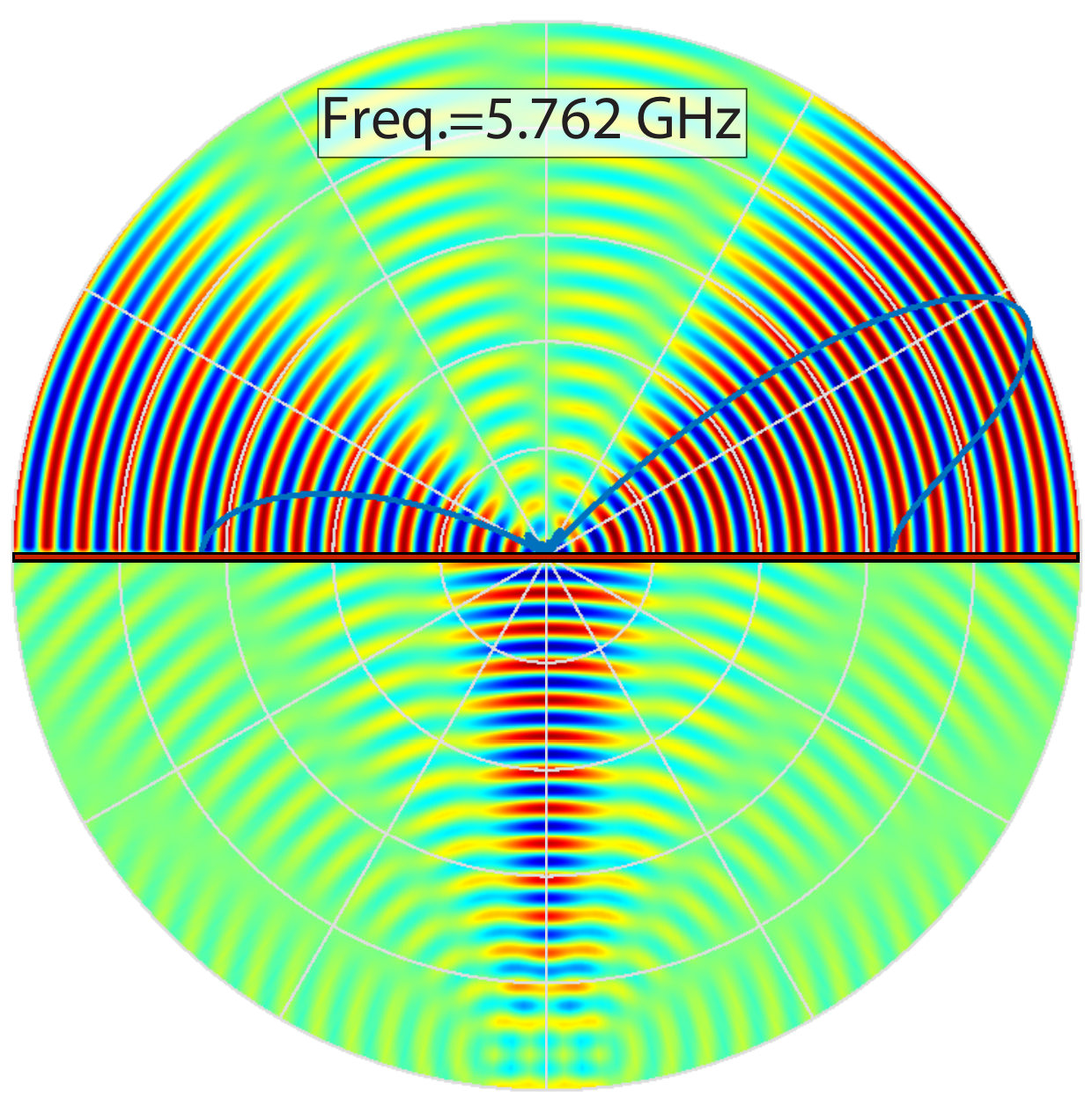}}	
		\subfigure[]{\label{fig:s2}
			\includegraphics[width=0.32\columnwidth]{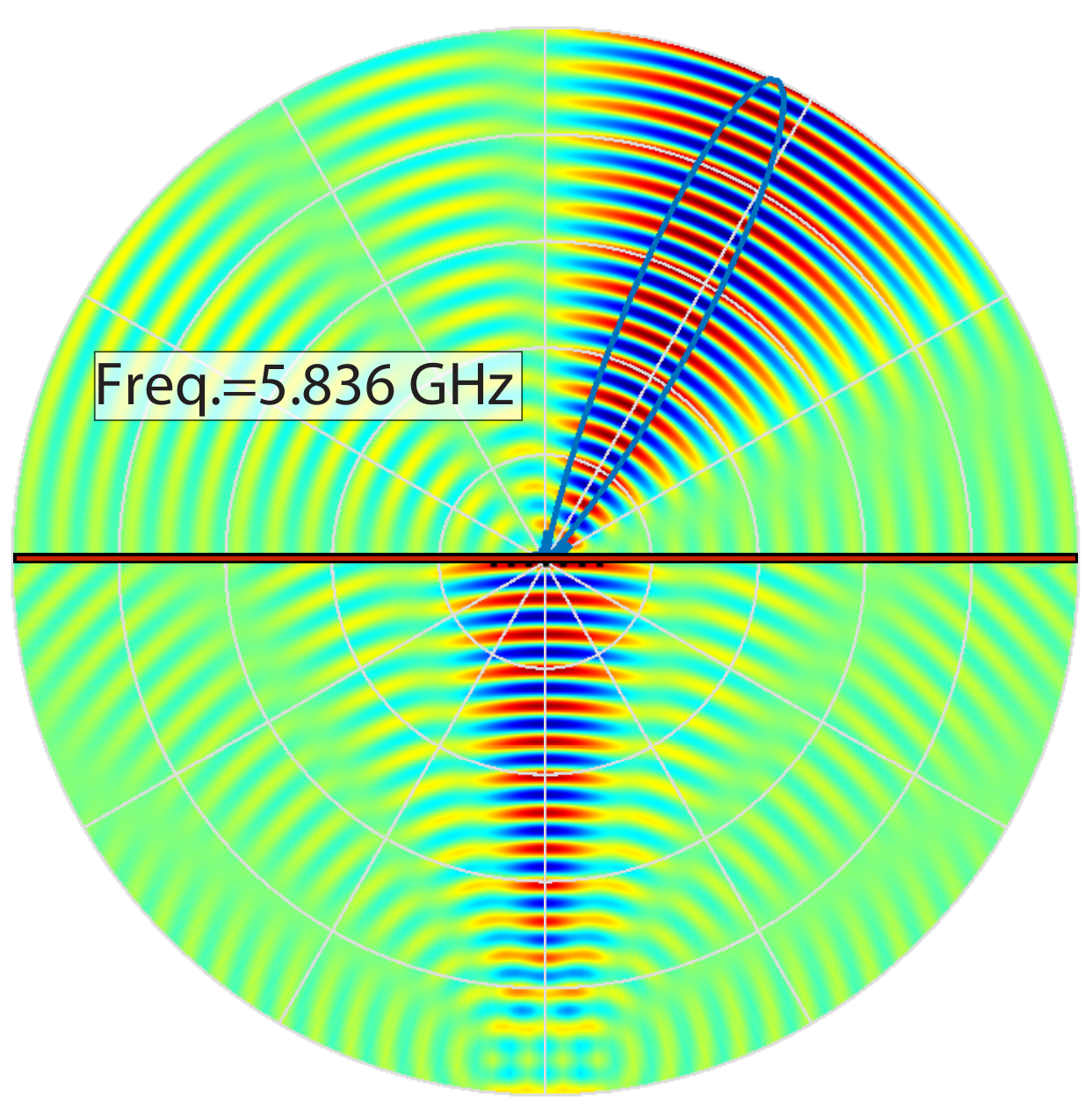}}	
		\subfigure[]{\label{fig:s3}
			\includegraphics[width=0.32\columnwidth]{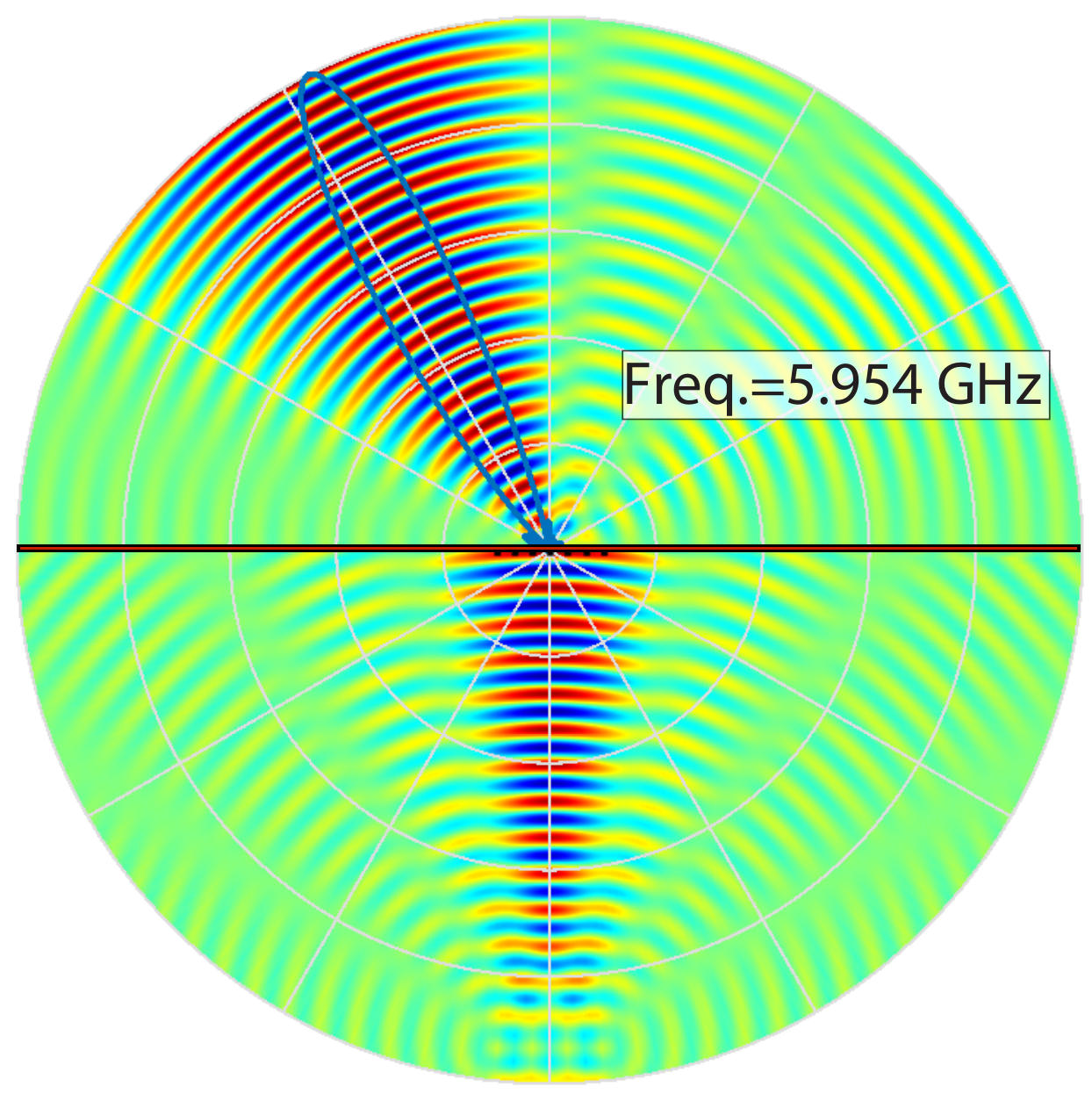}}
		\subfigure[]{\label{fig:BS03}
			\includegraphics[width=0.32\columnwidth]{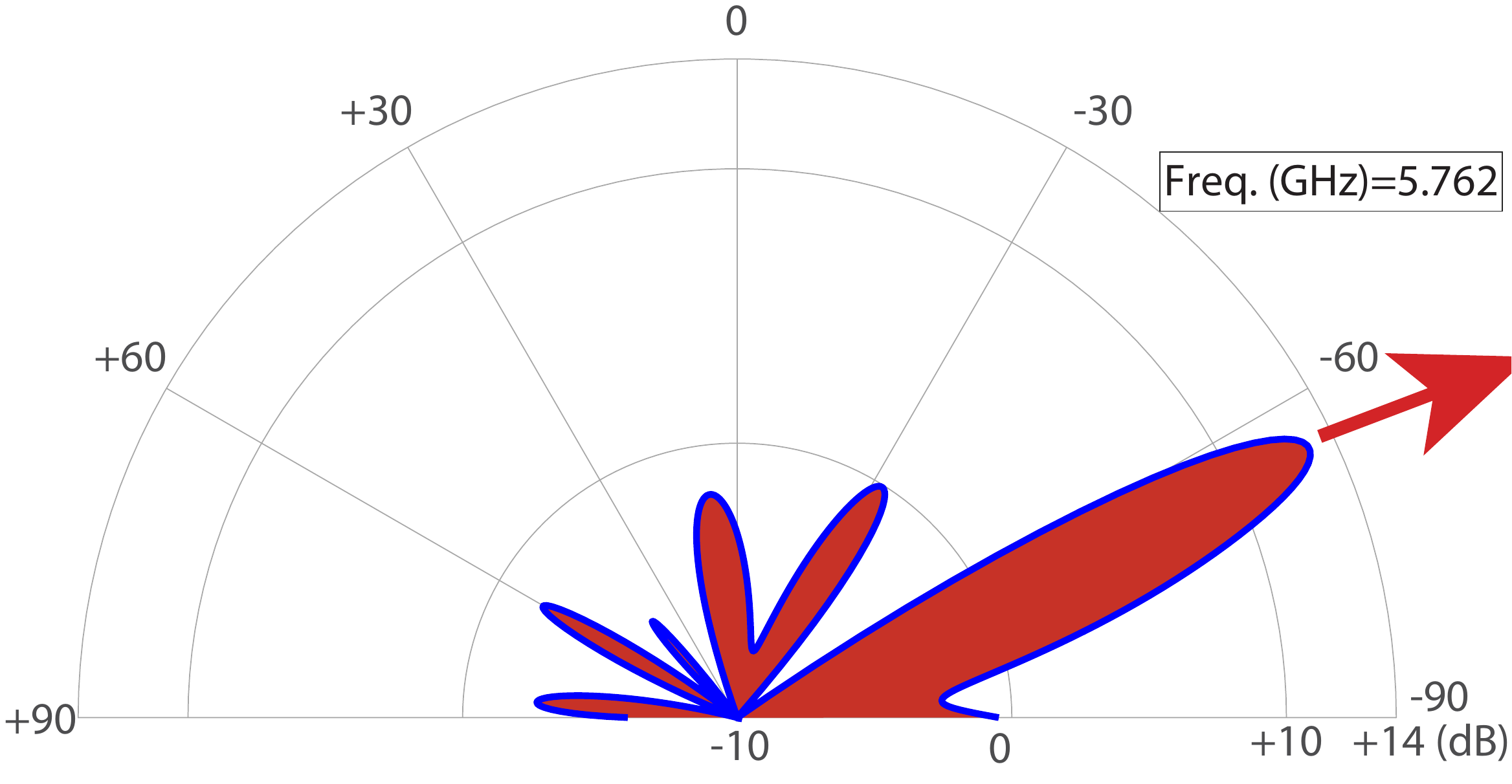}}	
		\subfigure[]{\label{fig:BS07}
			\includegraphics[width=0.32\columnwidth]{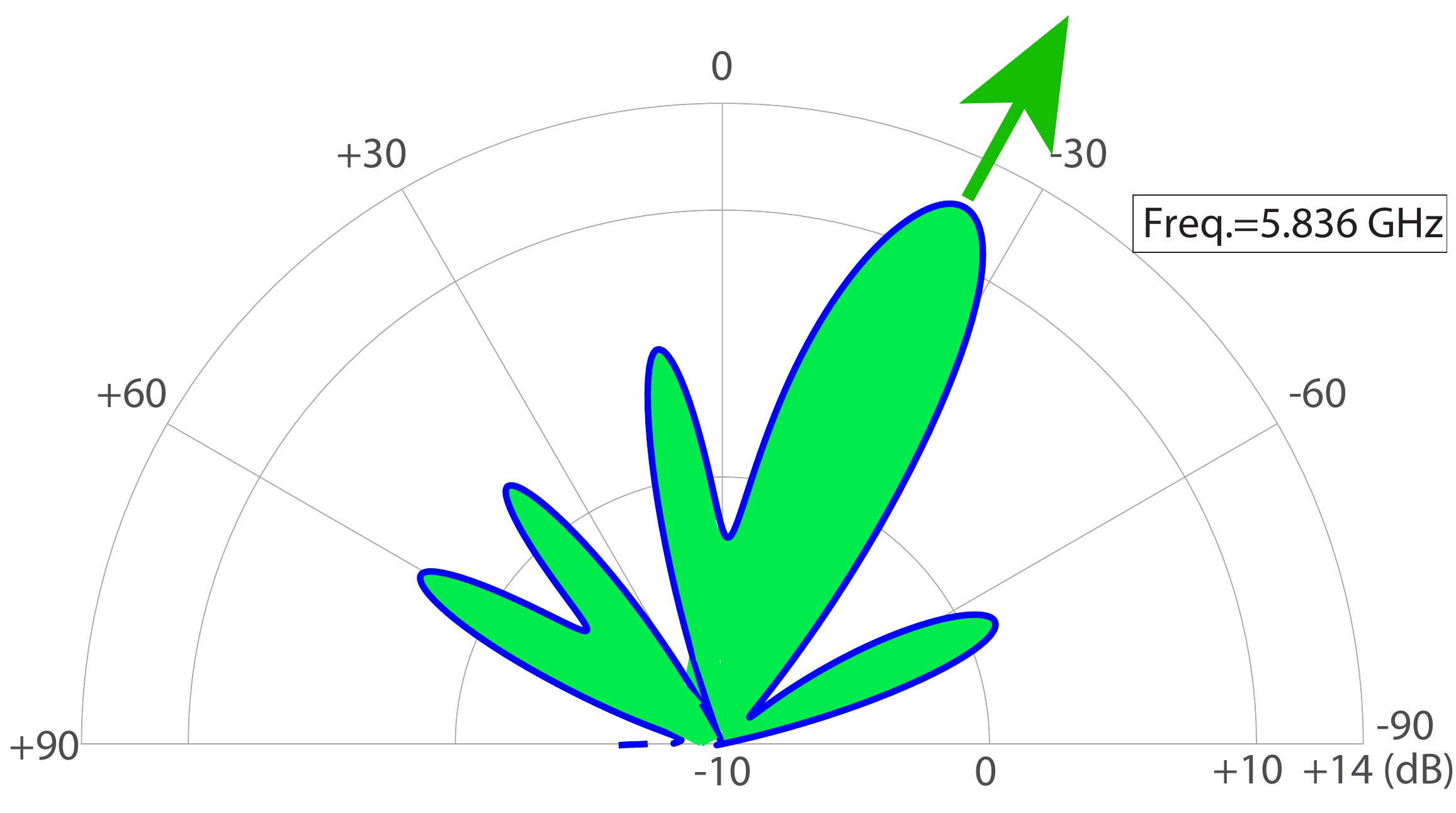}}
		\subfigure[]{\label{fig:BS9}
			\includegraphics[width=0.32\columnwidth]{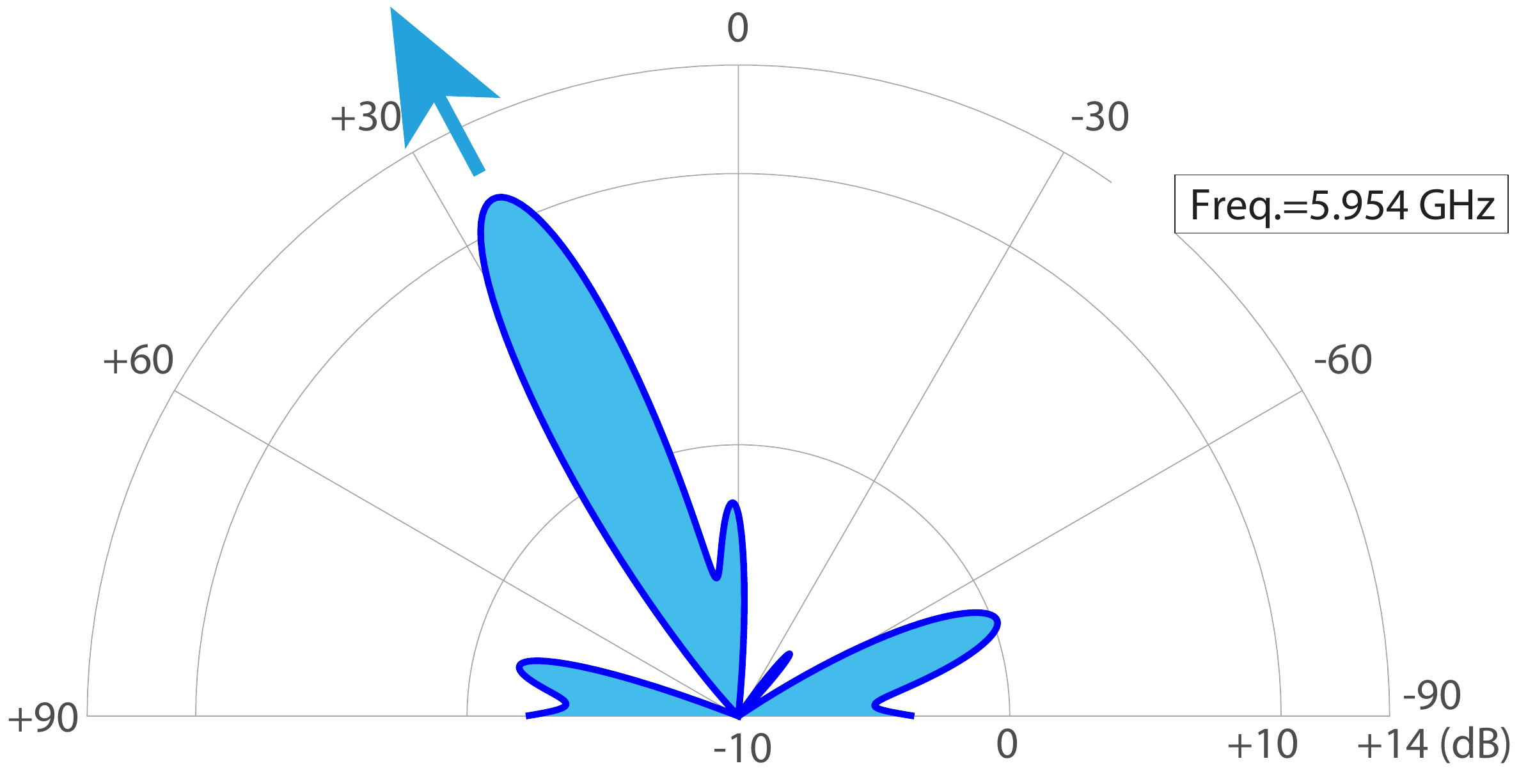}}
		\subfigure[]{\label{fig:BS04}
			\includegraphics[width=0.31\columnwidth]{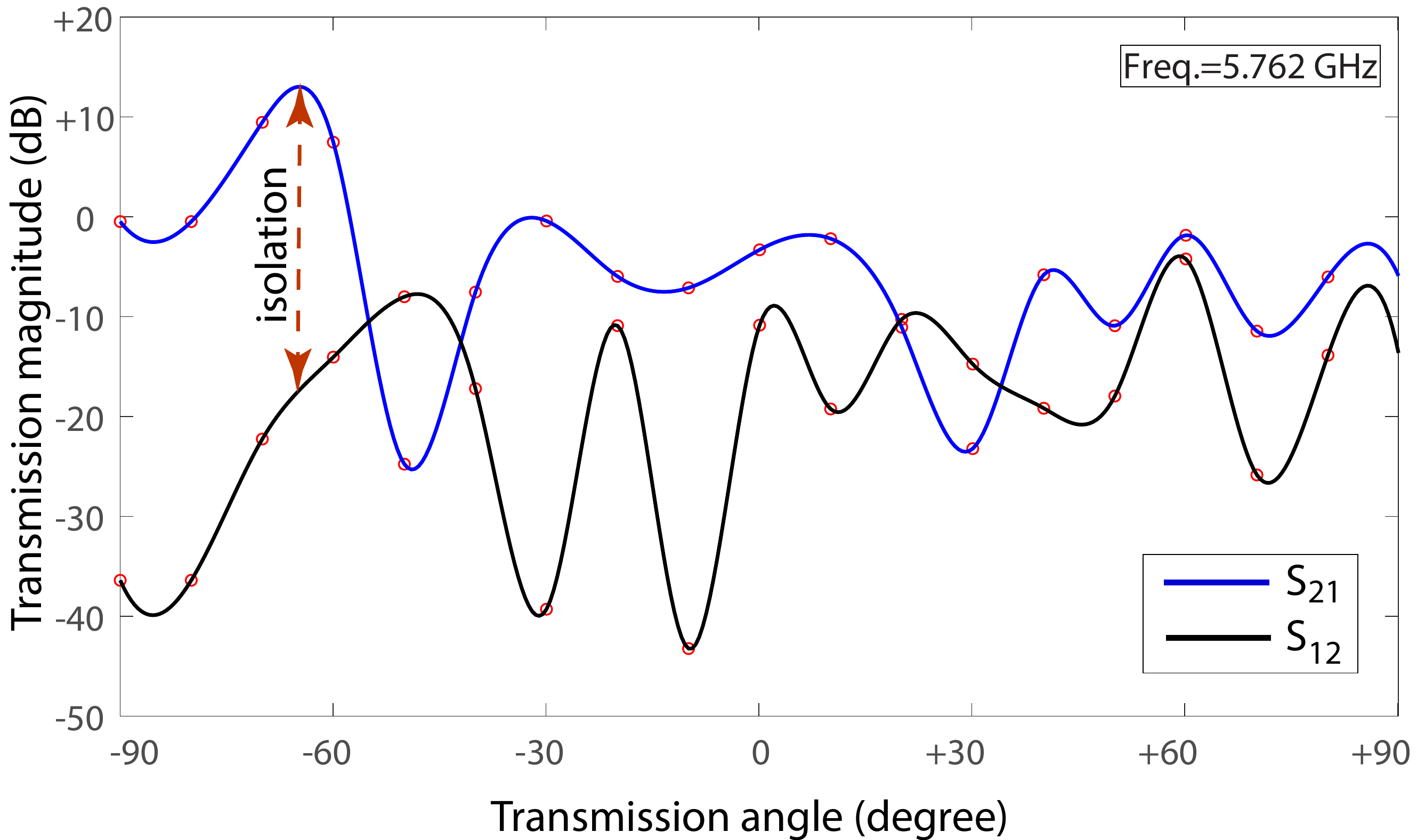}}	
		\subfigure[]{\label{fig:BS08}
			\includegraphics[width=0.31\columnwidth]{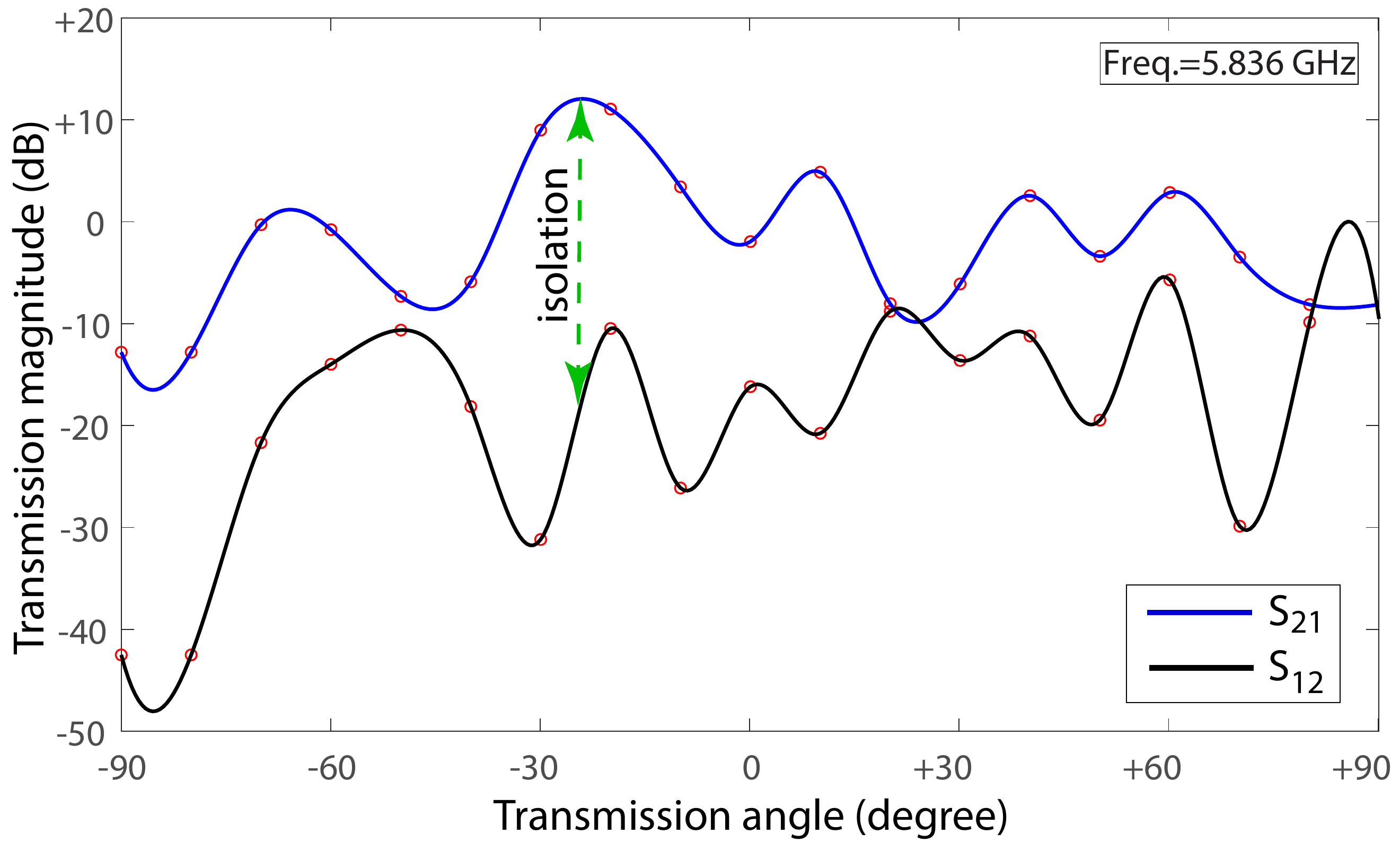}}
		\subfigure[]{\label{fig:BS10}
			\includegraphics[width=0.31\columnwidth]{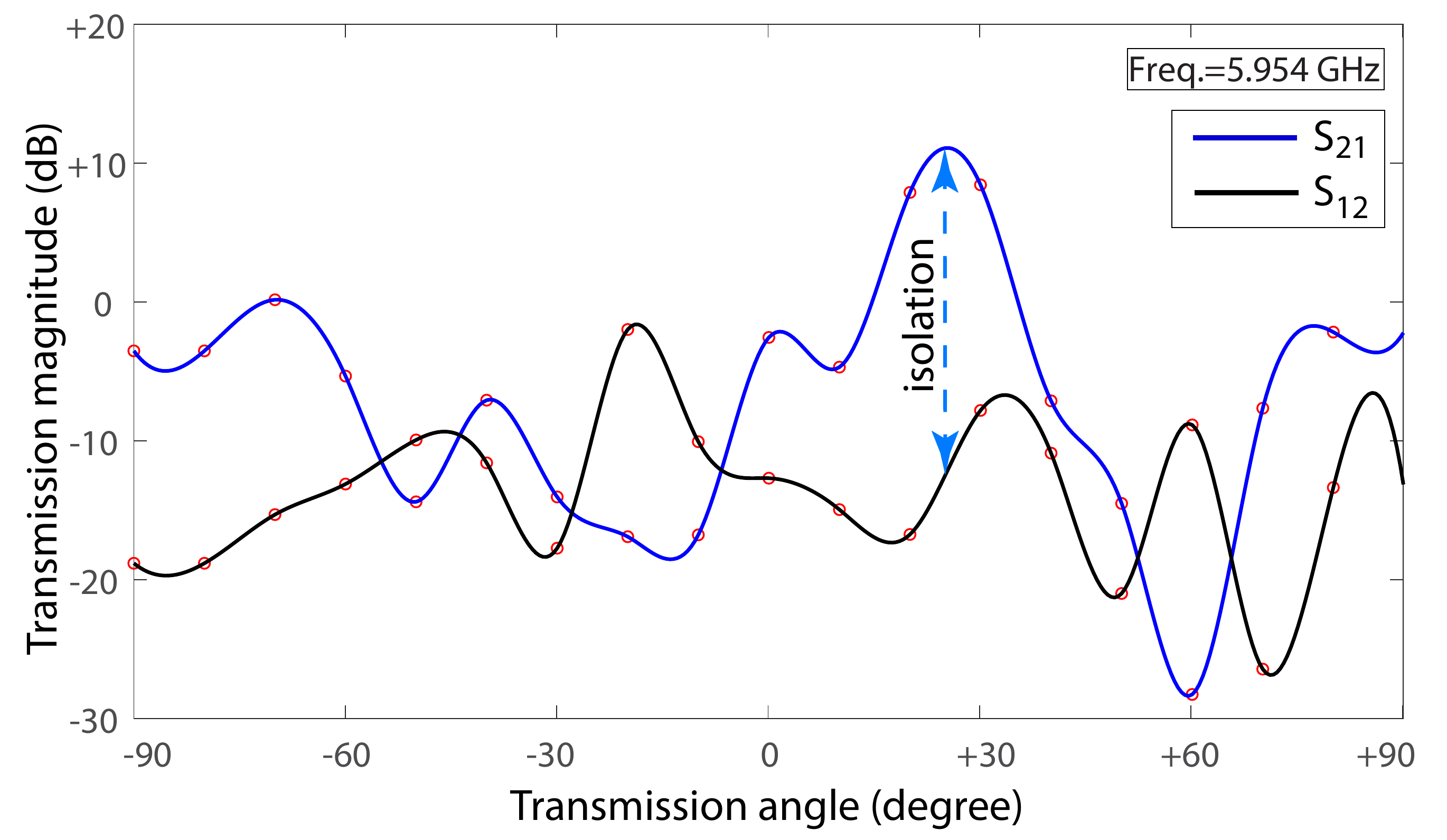}}
		\caption{Full-wave simulation and experimental results for nonreciprocal spatial decomposition of frequency components of the incident wave at 5.762 GHz, 5.836 GHz and 5.954 GHz. (a)-(c) Full-wave simulation results for electric field distribution. (d)-(i) Experimental results for forward and backward transmissions through the metasurface, showing a nonreciprocal prism-like functionality.}
		\label{fig:BS1}
	\end{center}
\end{figure*}

In the next experiment, we evaluate the far-field performance of the metasurface prism. Figure~\ref{fig:meas_su} shows an image of the far-field measurement set-up. Here, we fix the position of one horn antenna normal to the metasurface and rotate the other antenna from $0$ to $180^\circ$ with respect to the metasurface. Figures~\ref{fig:s1},~\ref{fig:s2} and~\ref{fig:s3} provide the full-wave simulation results of the electric field distribution for spatial decomposition of frequency components and prism functionality of the metasurface. Figures~\ref{fig:BS03} to~\ref{fig:BS10} plot the experimental results at different frequencies, showing nonreciprocal prism-like spatial decomposition of frequency components of the incident polychromatic wave comprising  three different frequency components at 5.762 GHz, 5.836 GHz and 5.954 GHz. Experimental results at three different frequencies show that at 5.762 GHz the maximum forward transmission occurs at $-65.4$ degrees with more than $12$ dB forward transmission gain and more than $28$ dB isolation between the forward and backward transmissions. In addition, at 5.836 GHz the maximum forward transmission occurs at $-24.2$ degrees with more than $11.54$ dB forward transmission gain and more than $27.3$ dB isolation between the forward and backward transmissions, and at 5.954 GHz the maximum forward transmission occurs at $+25.1$ degrees with more than $10.4$ dB forward transmission gain and more than $22.1$ dB isolation between the forward and backward transmissions.


\section{Discussion}

The experimental results show that the proposed nonreciprocal metasurface prism operates as expected, with remarkable efficiency. It exhibits the following additional favorable features. In addition to the prism-like spatial decomposition of frequency components, and a strong isolation between the forward and backward transmissions, the metasurface prism provides a desirable gain, which makes it particularly efficient for different wireless telecommunication applications as well as optical systems. Furthermore, in contrast to other nonreciprocal metasurfaces, the metasurface prism is polychromatic since patch antennas are fairly broadband and their bandwidth can be further enhanced by various standard band-broadening techniques~\cite{Garg_2001}. We shall stress that the metasurface prism presented here has not been optimized in this sense but already provides more than $4\%$ fractional bandwidth.

\section{Materials and methods}

The metasurface was realized using multilayer circuit technology, where two $7$~in $\times$~$7$~in RO4350 substrates with thickness $h = 30$~mil are assembled to realize a three metalization layer structure. The permittivity of the substrate is $\epsilon=\epsilon_\text{r}(1-j \tan \delta)$, with $\epsilon_\text{r} = 3.66$ and $\tan \delta = 0.0037$ at 10 GHz. The middle conductor of the structure (shown in Fig.~\ref{fig:arch}(a)) supports the DC feeding network and also acts as the RF ground plane for the patch antennas and transmission lines. Each side of the metasurface includes 16 microstrip patches, where the dimensions of the 2$\times$16 microstrip patches are $1.08$~in $\times$~$0.49$~in. The connections between the conductor layers are provided by an array of circular metalized via holes, where 32 vias of 45~mil diameter connect the DC bias network to the amplifiers, while the ground reference for the amplifiers is ensured by 32~sets of 6~vias of 30 mils diameter with 70~mils spacing. Furthermore, the RF path connection between the two sides of the metasurface is provided by 16 via holes, with optimized dimensions of $80$~mils for the via diameters, $120$~mils for the pad diameters and $210$~mils for the hole diameter in the via middle conductor. For the unilateral transistor-based amplifiers, we utilized 32 Mini-Circuits Gali-2+ Darlington pair amplifiers. Here, $C_\text{dcp}=$3~pF and $C_\text{out}=$3~pF, and $C_\text{bp1}=4.7$~pF, $C_\text{bp2}=1$~nF and $C_\text{bp3}=1$~uF.

\section*{Acknowledgements}

This work was supported in part by TandemLaunch Inc. and LATYS, Montreal, QC, Canada, and in part by the Natural Sciences and Engineering Research Council of Canada (NSERC). The authors would like to especially thank Mr. Gursimran Singh Sethi, Co-founder and Technical Leader of LATYS, and Dr. Omar Zahr, Director of Technology at TandemLaunch Inc., for their great help and support.

\section*{Author contributions statement}

S.T. carried out the analytical modeling, numerical simulations, sample fabrication, and measurements. G.V.E. conceived the idea, planned, coordinated, and supervised the work. All authors discussed the theoretical and numerical aspects and interpreted the results. All authors contributed to the preparation and writing of the manuscript. Correspondence and requests for materials should be addressed to Sajjad Taravati~(email: sajjad.taravati@utoronto.ca).

\section*{Additional information}
The authors declare no competing interests.

\bibliography{Taravati_Reference}

\end{document}